\documentclass[fleqn,usenatbib]{mnras}

\usepackage{newtxtext,newtxmath}
\usepackage[nolist]{acronym}

\usepackage[T1]{fontenc}
\usepackage{ae,aecompl}
\usepackage{graphicx}	

\usepackage{amsmath}	
\usepackage{amssymb}	

\usepackage{color}
\usepackage{rotating,lscape}

\usepackage[nolist]{acronym}
\usepackage{subfig} 

\newcommand{\g}[1]{\mbox {G1.9+0.3}}
\def\HI{\hbox{H\,{\textsc i}}}

\def\p0{\phantom{0}}

\newcommand{\D}{$^\circ$}
\def\arcmin{\hbox{$^\prime$}}
\def\arcsec{\hbox{$^{\prime\prime}$}}

\usepackage{xcolor}

\newcommand{\change}[1]{#1}

\begin{acronym}
\acro{AGN}{Active Galactic Nuclei}
\acro{ALMA}{Atacama Large Millimeter Array}
\acro{ATCA}{Australia Telescope Compact Array}
\acro{ATNF}{Australia Telescope National Facility}
\acro{ATOA}{Australia Telescope Online Archive}
\acro{AT20G}{Australia Telescope 20 GHz Survey}
\acro{ASKAP}{Australian Square Kilometre Array Pathfinder}
\acro{CASS}{CSIRO Astronomy and Space Science}
\acro{CABB}{Compact Array Broadband Backend}
\acro{CMZ}{Central Molecular Zone}
\acro{CSIRO}{Australian Commonwealth Scientific and Industrial Research Organisation}
\acro{CC}{Core Collapse}
\acro{CR}{Cosmic Ray}
\acro{CSS}{Compact Steep Spectrum}
\acro{CTA}{Cherenkov Telescope Array}
\acro{DM}{Dispersion Measure}
\acro{EMU}{Evolutionary Map of the Universe}
\acro{ESP}{Early Science Project}
\acro{FWHM}{Full Width at Half-Maximum power}
\acro{GC}{Galactic Center}
\acro{GLEAM}{GaLactic and Extragalactic All-sky MWA survey}
\acro{GPS}{Gigahertz Peak Spectrum}
\acro{HFP}{High Frequency Peaker}
\acro{ISM}{Interstellar Medium}
\acro{LMC}{Large Magellanic Cloud}
\acro{MCs}{Small \& Large Magellanic Clouds}
\acro{MCELS}{Magellanic Cloud Emission Line Survey}
\acro{MIRIAD}[{\tt MIRIAD}]{Multichannel Image Reconstruction, Image Analysis and Display}
\acro{MOST}{Molonglo Observatory Synthesis Telescope}
\acro{MRC}{Molonglo Reference Catalogue of Radio Sources}
\acro{MWA}{Murchison Widefield Array}
\acro{MW}{Milky Way}
\acro{NRAO}{National Radio Astronomy Observatory}
\acro{NVSS}{NRAO VLA Sky Survey}
\acro{OPAL}{Online Proposal Applications \& Links}
\acro{pc}{parsec\acroextra{: 1 pc $\simeq 3.09 \times 10^{16}$ m}}
\acro{PMN}{Parkes-MIT-NRAO}
\acro{PN}{Planetary Nebula}
\acro{PWN}{Pulsar Wind Nebulae}
\acro{PNe}{Planetary Nebulae}
\acro{POSSUM}{Polarisation Sky Survey. of the Universe's Magnetism}
\acro{RM}{Rotation Measure}
\acro{QSO}{Quasi-Stellar Objects}
\acro{RFI}{Radio-Frequency Interference}
\acro{RMS}{Root Mean Squared}
\acro{RM}{Rotation Measure}
\acro{RSG}{Red Super Giant}
\acro{SCEM}{School of Computing, Engineering and Mathematics}
\acro{SED}{Spectral Energy Distribution}
\acro{SFG}{Star Forming Galaxies}
\acro{SI}[$\alpha$]{Spectral Index\acroextra{, $S \propto \nu^\alpha$}}
\acro{SKA}{Square Kilometre Array}
\acro{SMB}{Super Massive Black Hole}
\acro{SMC}{Small Magellanic Cloud}
\acro{SN}{Supernova}
\acro{SNRs}{Supernova Remnants}
\acro{SNR}{Supernova Remnant}
\acro{SUMSS}{Sydney University Molonglo Sky Survey}
\acro{WISE}{Wide-Field Infrared Survey Explorer}
\acro{VLA}{Very Large Array} 
\acro{VLBI}{Very Long Baseline Interferometry} 
\acro{MYSO}{Massive Young Stellar Object}
\acro{HFPs}{High Frequency Peakers}
\acro{ESP}{Early Science Project}
\acro{YSOs}{Young Stellar Objects}
\acro{IRAC}{Infrared Array Camera}
\acro{BL-Lac}{BL~Lacertae}
\acro{FSRQs}{Flat Spectrum Radio Quasars}
\acro{IR}{Infrared}
\end{acronym}

\title[SNR \g1 Radio Observations]{Radio Observations of Supernova Remnant \g1}

\author[K. Luken et al.]{Kieran J. Luken,$^{1,2}$\thanks{E-mail:k.luken@westernsydney.edu.au}
Miroslav D. Filipovi\'c,$^{1}$
Nigel I. Maxted,$^{3,1}$
Roland Kothes,$^{4}$
\newauthor
Ray P. Norris,$^{1,2}$
James R. Allison,$^{5}$
Rebecca Blackwell,$^{6}$
Catherine Braiding,$^{7,6}$
\newauthor
Robert Brose,$^{8,9}$
Michael Burton,$^{10}$
Ain Y. De Horta,$^{1}$
Tim J. Galvin,$^{1,11}$
\newauthor
Lisa Harvey-Smith,$^{7,1}$
Natasha Hurley-Walker,$^{12}$
Denis Leahy$^{13}$
Nicholas O. Ralph,$^{1}$
\newauthor
Quentin Roper,$^{1}$
Gavin Rowell,$^{6}$
Iurii Sushch,$^{9,14,15}$
Dejan Uro\v sevi\' c$^{16,17}$
\newauthor
and Graeme F. Wong$^{18,1,7}$
\\
$^{1}$Western Sydney University, Locked Bag 1797, Penrith, NSW 2751, Australia\\
$^{2}$CSIRO Astronomy and Space Sciences, Australia Telescope National Facility, PO Box 76, Epping, NSW 1710, Australia\\
$^{3}$School of Science, The University of New South Wales, Australian Defence Force Academy, Canberra, ACT 2600, Australia\\
$^{4}$Dominion Radio Astrophysical Observatory, Herzberg Programs in Astronomy and Astrophysics, National Research Council Canada,\\ PO Box 248, Penticton, BC V2A 6J9, Canada \\
$^{5}$Sub-Dept. of Astrophysics, Department of Physics, University of Oxford, Denys Wilkinson Building, Keble Rd., Oxford OX1 3RH, UK\\
$^{6}$School of Physical Sciences, The University of Adelaide, Adelaide 5005, Australia\\
$^{7}$School of Physics, The University of New South Wales, Sydney 2052, Australia\\
$^{8}$DESY, 15738 Zeuthen, Germany\\
$^{9}$Institute of Physics and Astronomy, University of Potsdam, 14476 Potsdam, Germany\\
$^{10}$Armagh Observatory and Planetarium, College Hill, Armagh, BT61 9DG, Northern Ireland, UK\\
$^{11}$CSIRO Astronomy and Space Science, PO Box 1130, Bentley WA 6102, Australia\\
$^{12}$International Centre for Radio Astronomy Research, Curtin University, Bentley, WA 6102, Australia\\
$^{13}$Department of Physics and Astronomy, University of Calgary, University of Calgary, Calgary, Alberta, T2N 1N4, Canada\\
$^{14}$Centre for Space Research, North-West University, 2520 Potcheftroom, South Africa\\
$^{15}$Astronomical Observatory of Ivan Franko National University of L'viv, vul. Kyryla i Methodia, 8, L'viv 79005, Ukraine\\
$^{16}$Department of Astronomy, Faculty of Mathematics, University of Belgrade, Studentski trg 16, 11000 Belgrade, Serbia\\
$^{17}$Isaac Newton Institute of Chile, Yugoslavia Branch\\
$^{18}$Pawsey Supercomputing Centre, 26 Dick Perry Ave, Kensington 6151, WA, Australia\\
}

\date{Accepted XXX. Received YYY; in original form ZZZ}

\pubyear{2019}

\begin{document}
\label{firstpage}
\pagerange{\pageref{firstpage}--\pageref{lastpage}}
\maketitle

\begin{abstract}
We present 1 to 10\,GHz radio continuum flux density, spectral index, polarisation and \ac{RM} images of the youngest known Galactic \ac{SNR} \g1, using observations from the \ac{ATCA}. We have conducted an expansion study spanning 8 epochs between 1984 and 2017, yielding results consistent with previous expansion studies of \g1. We find a mean radio continuum expansion rate of ($0.78 \pm 0.09$)~per~cent~year$^{-1}$ (or $\sim8900$~km~s$^{-1}$ at an assumed distance of 8.5~kpc), although the expansion rate varies across the \ac{SNR} perimeter. In the case of the most recent epoch between 2016 and 2017, we observe faster-than-expected expansion of \change{the northern region}. We find a global spectral index for \g1\ of $-0.81\pm0.02$ (76\,MHz--10\,GHz). Towards the northern region, however, the radio spectrum is observed to steepen significantly ($\sim -$1). Towards the two so called (east \& west) ``ears'' of \g1, we find very different \ac{RM} values of 400-600~rad~m$^{2}$ and 100-200~rad~m$^{2}$ respectively. The fractional polarisation of the radio continuum emission reaches (19 $\pm$ 2)~per~cent, consistent with other, slightly older, \acp{SNR} such as Cas~A.
\end{abstract}




\acresetall
\section{Introduction}
 \label{s:intro}

There are currently only $\sim$10 confirmed `young' (defined as being less than 2000 years old) Galactic \acp{SNR} out of a predicted $\sim$50  \citep{1991ARA&A..29..363V,2003astro.ph.10859C}. The \ac{SNR} \g1 is believed to be the youngest in the \ac{MW} with an age (calculated from its expansion rate) of $\sim$150 years \citep{2017ApJ...837L...7B,2014SerAJ.189...41D,2009ApJ...695L.149R,2008ApJ...680L..41R,2008MNRAS.387L..54G,2011ApJ...737L..22C}. \citet{2017MNRAS.468.1616P} estimated the age of \g1 as $\sim$120 years, based on an analysis of the hydrodynamical and radio evolution of this young \ac{SNR}. Previous studies suggest \g1\ is a Type~Ia \ac{SNR} \citep{2013ApJ...771L...9B}. \change{The high expansion velocity} of \g1, absence of an obvious \ac{PWN}, and bilateral symmetry of the X-ray emission have all been previously used as evidence \citep{2017ApJ...837L...7B}. \citet{2013ApJ...771L...9B} further postulated that only a very unusual core-collapse event could reproduce the observations, while a reasonable \ac{SN} Type~Ia model can reach the observed size and velocity with a mean external density of $\sim$0.02~cm$^{−3}$ \citep{2010ApJ...724L.161B}. Therefore, detailed studies of this fast evolving \ac{SNR} will give us unprecedented insight into the evolution of \acp{SNR} in general, with a particular interest in the early stages of their evolution, the dynamics of \ac{SN} ejecta and on particle acceleration. 

Upon its discovery with the \acl{VLA} \citep[\acs{VLA}; \acused{VLA}][]{Green:1984}, \g1 was noted to have a radio brightness comparable to the Tycho and Kepler SNRs with a spectral index of $\alpha\sim-0.6$\footnote{Spectral index is defined as $S\propto\nu^\alpha$}. \ac{MOST} Galactic Survey data resolved a shell-like morphology with diameter 1\farcm2 \citep{Gray:1994}. Subsequent studies of 20/90\,cm \ac{VLA} data characterised the \ac{SNR} as having diameter $\lesssim$1\arcmin\ and spectral index of $\alpha=-0.93\pm0.23$ \citep{LaRosa:2000,Nord:2004}, with the steep spectral index suggesting the radio emission is primarily synchrotron based. \citet{Farnes:2012PhD} later mapped the spatial variation in polarisation and spectral index, noting flatter spectra in the NW and SE of the remnant. 

Distance is a key variable for calculating expansion velocity and age. \citet{Nord:2004} inferred the distance to be $<$7.8~kpc using the lack of 74~MHz absorption by the Galactic Plane as an indicator that \g1\ is on the near side of the \ac{GC}. This distance is consistent with X-ray absorption studies \citep{2008MNRAS.387L..54G}. Subsequently, \HI\ absorption in the so-called `Feature-I' gas structure \citep[See Section~\ref{section:absorption}]{1975MNRAS.171..659C} was observed in the \g1 radio continuum emission, implying that the \ac{SNR} is beyond Feature-I \citep{2014IAUS..296..197R}. Feature-I extends $>$5$^{\circ}$ in Galactic longitude and does not appear in \HI-absorption towards the Sagittarius A$^*$ radio continuum, so this component lies beyond the precise \ac{GC} distance. It follows that \g1, as it corresponds with Feature-I, must also lie beyond the \ac{GC}\change{, which, following IAU standards, we assume to lie at a distance of 8.5~kpc \citep{1986MNRAS.221.1023K}. However, we do note the recent disagreement in this fundamental parameter by \citet[$\sim$7.4~kpc; ][]{2014MNRAS.441.1105F} and \citet[$\sim$8.3~kpc; ][]{2016ApJS..227....5D}. } The far 3\,kpc-Expanding arm at line of sight velocity $\sim +$50\,km~s$^{-1}$, does not appear in absorption \citep{2014IAUS..296..197R}, therefore \g1 is in front of this component. \change{We therefore assume \g1 to have a distance of 8.5\,kpc, which is consistent with both X-ray absorption studies \citep{2011ApJ...737L..22C}, and the lower and upper distance constraints derived from Feature-I and the far Expanding arm, respectively. }

\citet{2008MNRAS.387L..54G} re-observed \g1 at 4.86~GHz using the \ac{VLA} after \citet{2008ApJ...680L..41R} used 2007 {\it Chandra} images to show \g1 had expanded significantly since 1985 and its X-ray emission appeared to be predominantly synchrotron in nature. By comparing these new VLA observations with the 1985 VLA observations made at 1.49~GHz, \citet{2008MNRAS.387L..54G} determined that \g1 had expanded by 15$\pm$2~per~cent over 23 years ($\sim$0.65~per~cent per yr). Using only the 1985 and 1989 VLA observations, \citet{Gomez:2009} derived a smaller expansion rate of $0.46\pm0.11$~per~cent, and estimated the \g1 age to be $220 ^{+70}_{-45}$ yrs. \citet{2014SerAJ.189...41D} later used all available \ac{ATCA} and VLA radio-continuum observations at 6\,cm, to estimate a median expansion rate of 0.563$\pm$0.078~per~cent per yr between 1984 and 2009. It was noted that the apparent expansion of \g1 was slower (0.484~per~cent~per~yr) in the 1980s compared to recent epochs (2014; 0.641~per~cent~per~yr).

X-ray observations have also been used to measure expansion, with \citet{2011ApJ...737L..22C} finding an expansion rate of 0.642~$\pm$0.049~per~cent~per~yr and a flux density increase of 1.7~$\pm$1.0~per~cent~per~yr by comparing 2007 and 2009 \textit{Chandra} images. A simple uniform-expansion model leads to a \g1 age estimate of 156$\pm$11\,years old assuming no deceleration, however, a uniform expansion model is probably an oversimplification. \citet{2013ApJ...771L...9B} recorded ejecta from \g1 to have speeds as large as $\sim$18\,000~km~s$^{-1}$. In their study, an implied abundance inhomogeneity of Fe-rich northern ejecta and Si/S-rich eastern ejecta were said to be consistent with asymmetrical Type~Ia \ac{SN} explosion models.

As \g1 expanded, it also brightened. \citet{2008MNRAS.389L..23M} found that the flux density of \g1 at 843~MHz increased by ($1.22\pm ^{0.24}_{0.16}$)~per~cent~per~yr over the last two decades. From simulations based on the non-linear diffuse shock acceleration (NLDSA) model, \citet{2017MNRAS.468.1616P} found that the radio flux density should have increased by $\sim1.8$~per~cent~per~yr over the past two decades. Such behaviour would be consistent with a \ac{SNR} sweeping up the surrounding \ac{ISM} increasing number of particles which can be injected into the NLDSA process, gaining ultra-relativistic energies and emitting synchrotron radiation. Additionally, this numerical model predicts that the radio flux density will increase, reaching its maximum value around 500~years from now. During the late free expansion phase, the \ac{SNR} flux density will start to decrease. The beginning of Sedov phase will start around 1700 years after the initial \ac{SN} explosion. \citet{2017MNRAS.468.1616P} emphasised that in this stage of evolution of \g1, we are witnessing the fastest increase in radio brightness this \ac{SNR} will ever produce. Moreover the steep radio continuum spectrum of \g1 obtained from various observations can be explained as a result of the efficient NLDSA accompanying with strong magnetic field amplification \citep{2017MNRAS.468.1616P}. 

However, \citet{2008ApJ...680L..41R} noted that the radii of the X-ray and radio shells of \g1 differ by $\approx 20$~per~cent. This difference can not be explained if electrons are accelerated only at the forward shock of \g1 as the maximum offset possible in the case of very efficient magnetic field amplification and thus efficient synchrotron cooling is of the order of 5~per~cent. \citet{2019arXiv190602725B} showed that the different expansion and brightening of the radio and X-ray shells can be explained if the X-ray emission originates at the forward shock and the radio emission mainly at the reverse shock. Efficient particle acceleration at the reverse shock has been proposed earlier \citep{2013A&A...552A.102T}. Combined with the young age of \g1, a reverse-shock density three times higher than the forward-shock density and a reverse shock speed of $\approx 5600\,$km~s$^{-1}$ in the plasma frame, \g1 is a prime candidate for the detectable non-thermal emission from the reverse shock.

\g1 has previously been observed with the \ac{ATCA}. However, the \ac{CABB} upgrade \citep{CABB} increased the resolution, sensitivity and bandwidth of the array. This allows us a detailed look at the \ac{SNR} morphology and spectrum, as well as providing additional epochs with which to measure the expansion of this young Galactic \ac{SNR}. Observations of \g1 from 1984, 1985, 1987, 1989 and 2008 (taken from the \ac{VLA} archives) and our \ac{ATCA} observations (from 2009, 2016 and 2017) give us an opportunity to conduct a high-precision study of its expansion in the radio regime.

In this paper, we present the results from 2016 \& 2017 ATCA observations of \g1. In Section~\ref{section:absorption} we investigate \HI\ absorption and CO(1-0) structure towards \g1 in the context of previous studies. \change{We derive the expansion rate of \g1\ using the new observations taken using the \ac{ATCA} and previous images from the \ac{ATCA} and \ac{VLA} in Section~\ref{section:expansion}}. This is followed by our polarisation and \ac{RM} studies of the \ac{SNR} in Section~\ref{section:polarisation}. In Section\,\ref{section:specIndex}, we present spectral index maps, and calculate a revised \g1 spectral index using our \ac{ATCA} observations and data from the \ac{GLEAM} project \citep{GLEAM:2015}. In Section\,\ref{section:brightening}, we then calculate the general and local radio flux density increase of \g1.

The results are discussed in Section~\ref{section:discussion}, with the conclusion presented in Section~\ref{section:conclusion}.

\section{Data and Observations}
 \label{section:Data Obs Methods}

We observed the \ac{SNR} \g1 in 2016 and 2017 using the \ac{ATCA} across three twelve hour periods (Project C1952; Table~\ref{table:obsDetails}). Using the EW352 array on 26-27 January 2016, we observed in spectral line (observed at 1421~MHz, 1610~MHz, 1664~MHz, 1666~MHz and 1719~MHz) and continuum mode (2100, 5000 and 9000~MHz) using a frequency switching technique. The 6B array was used on 8-9 March 2016 in continuum mode (2.1, 5 and 9~GHz). On 20-21 May 2017, we used the 6A array in spectral line mode (observing at 1421~MHz, 1610~MHz, 1664~MHz, 1666~MHz and 1719~MHz) and continuum mode (2100, 5000 and 9000~MHz) using a frequency switching mode. Over this time and across all arrays used, there was a total of 44 unique baselines, covering a range of spacings between 30~m and 5969~m and giving us excellent \textit{uv} coverage. 

The same flux density calibrator (1934-628) and phase calibrator (1710-269) were used for all observations. The complete observational details are available in Table \ref{table:obsDetails}. 

\begin{figure*}
\centering
\includegraphics[trim=0 0 0 0, width=.4\textwidth]{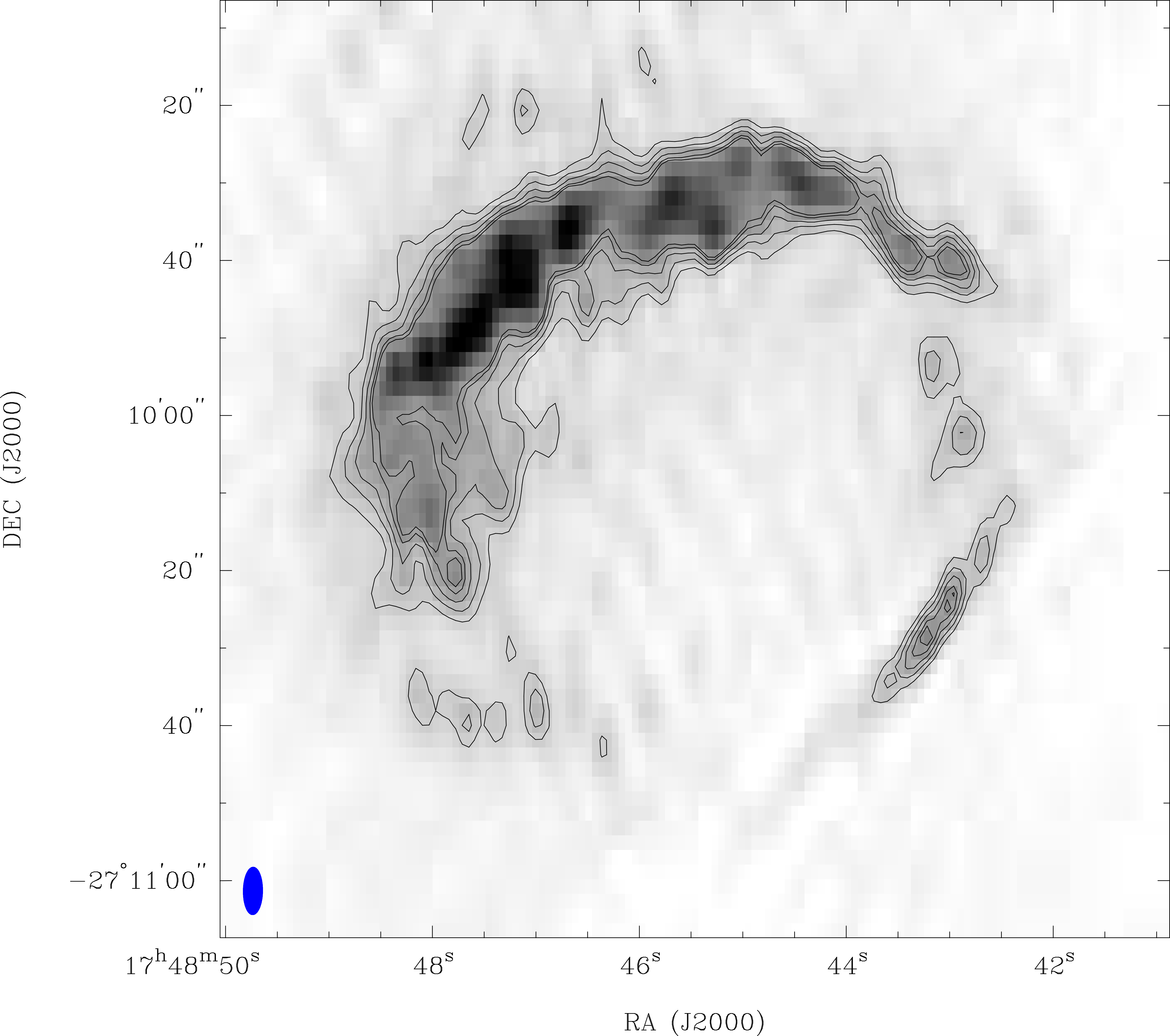}
\includegraphics[trim=0 0 0 0, width=.4\textwidth]{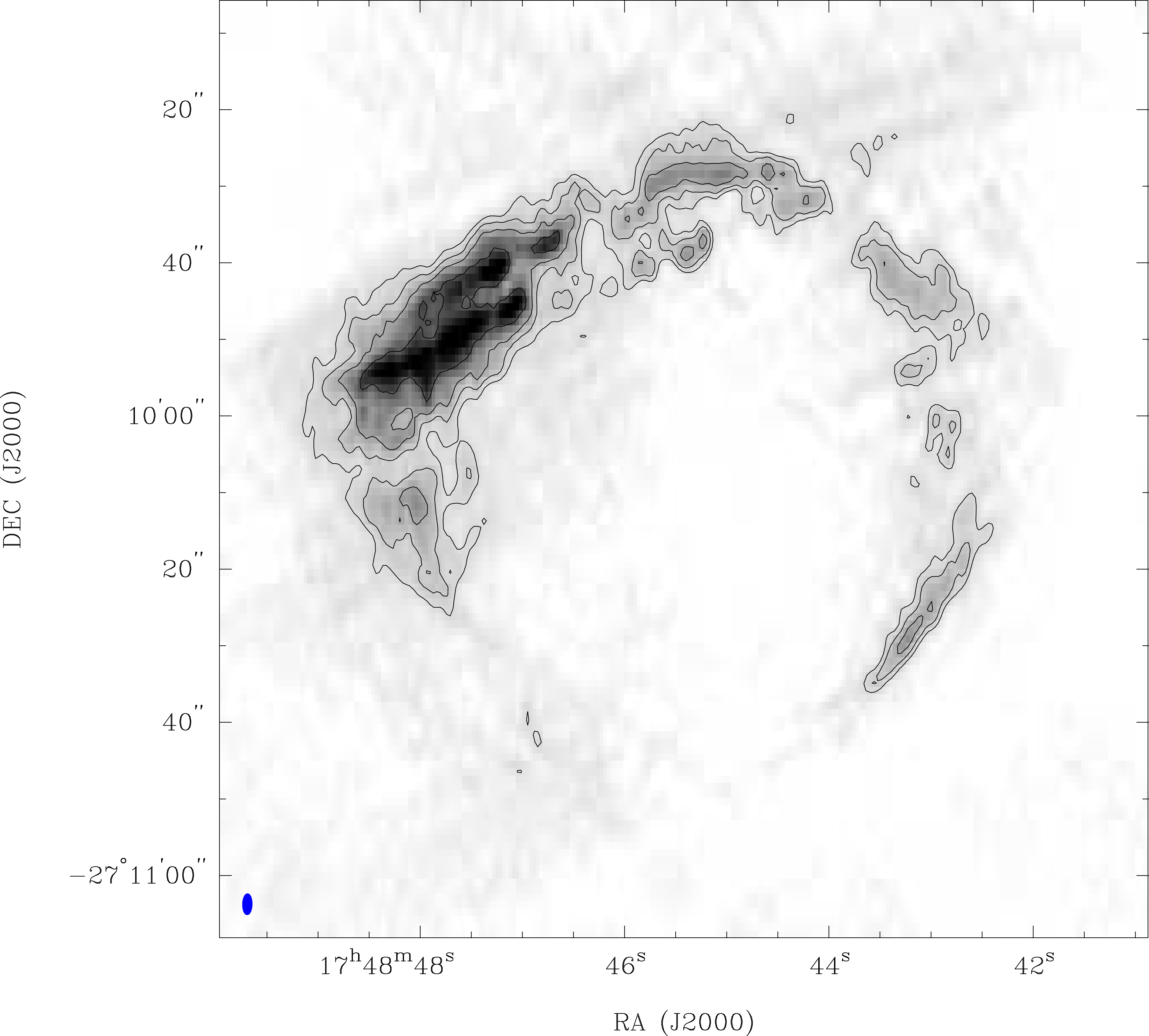}
\includegraphics[trim=0 0 0 0, width=.4\textwidth]{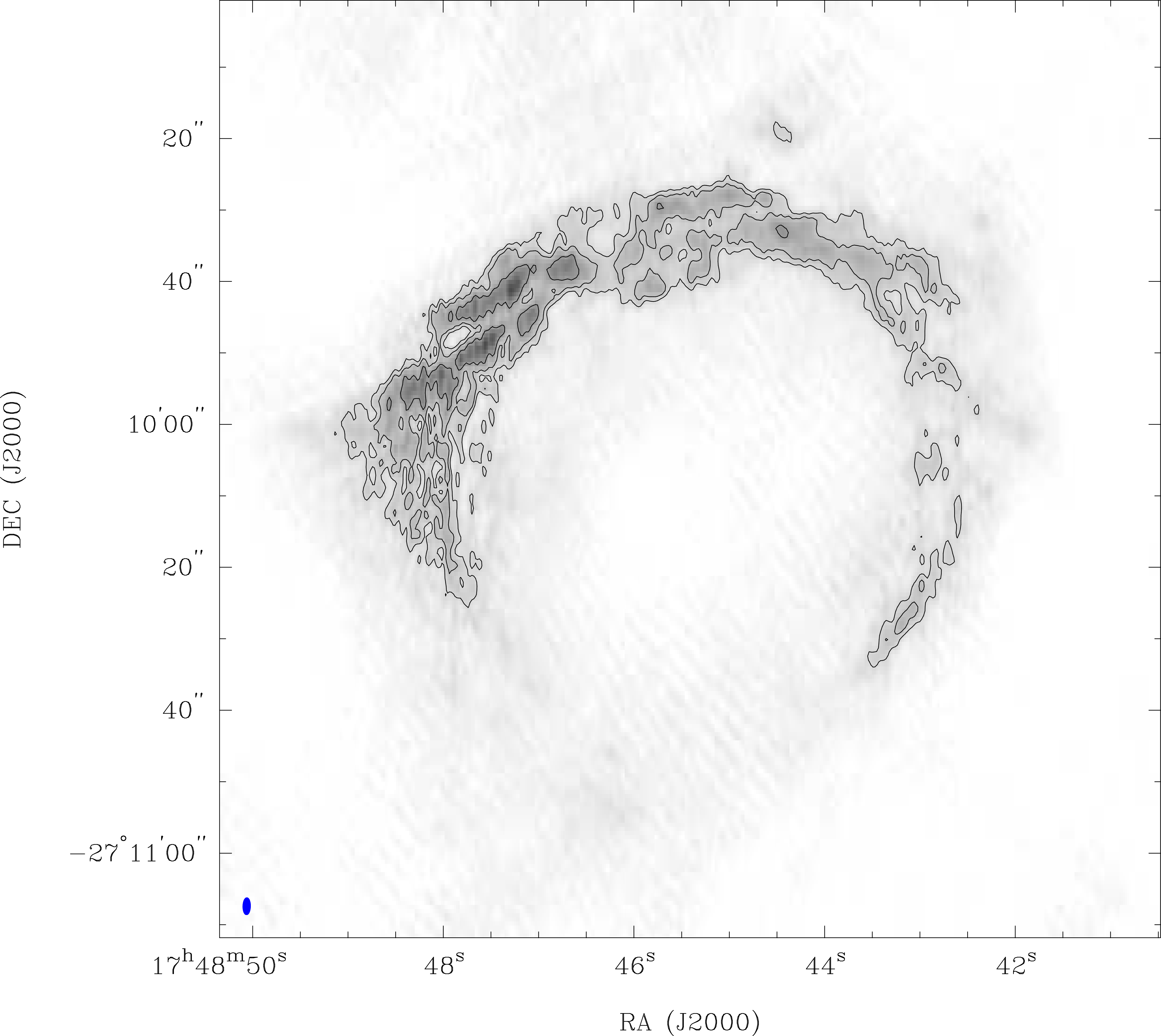}
\includegraphics[trim=0 0 0 0, width=.4\textwidth]{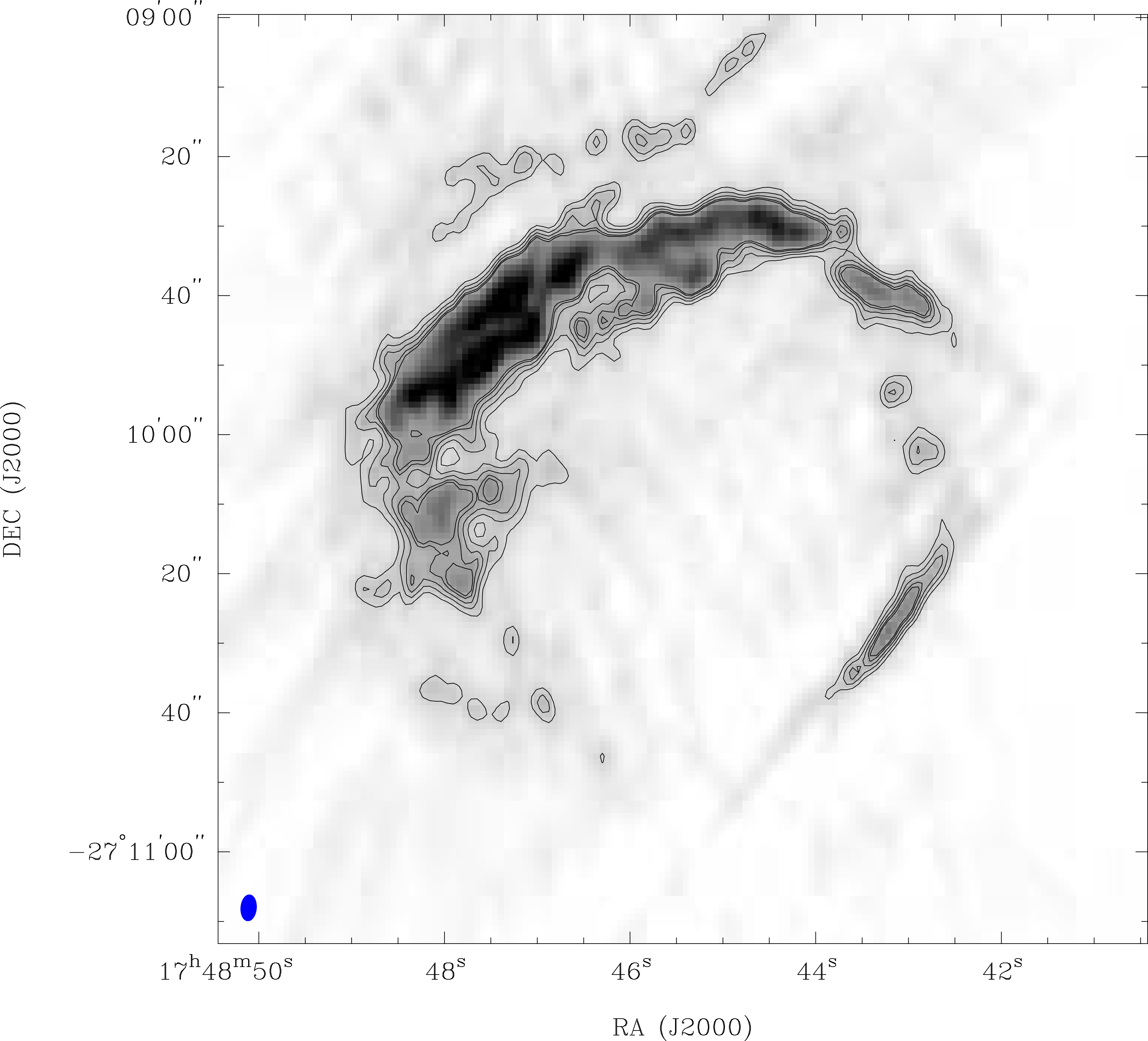}
\includegraphics[trim=0 0 0 0, width=.4\textwidth]{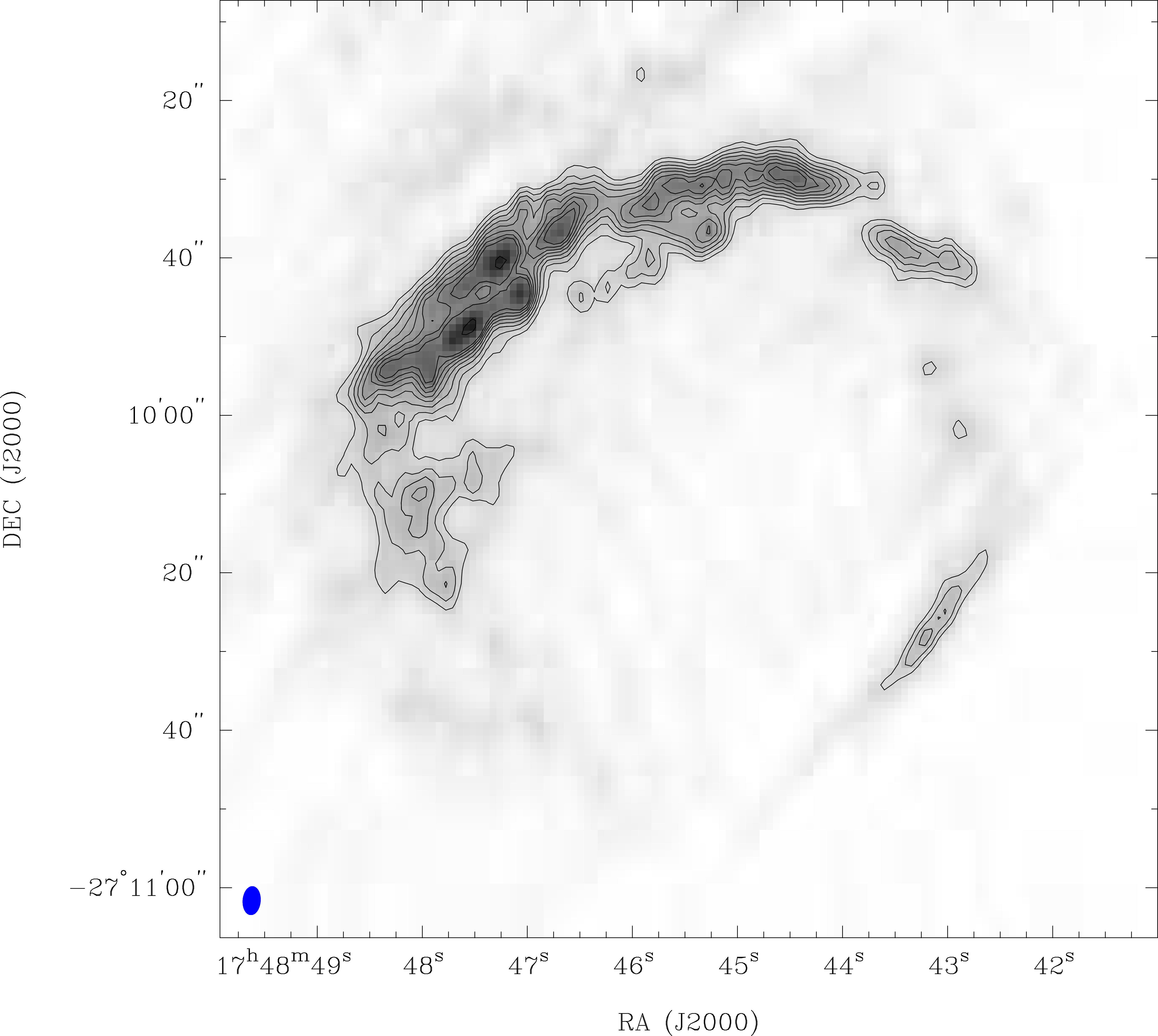}
\caption{Radio images produced from the 2016 observations of \g1\ using the \ac{ATCA}. From top-left to bottom: 2.1~GHz, 5~GHz, 9~GHz, 2.1 and 5~GHz combined and 2.1, 5 and 9~GHz combined. The ellipse in the lower-left corner of each is representative of the beam size (Details in Table \ref{table:contImageDetails}). Contour levels for the 2.1~GHz image are 15, 20, 25, 30 and 33~$\sigma$ ($\sigma = $~0.16~mJy~beam$^{-1}$). 5.0~GHz contours are 2, 3, 5 and 10~$\sigma$ ($\sigma = $~0.10~mJy~beam$^{-1}$). 9.0~GHz contours are 2, 3, 5 and 10~$\sigma$ ($\sigma = $~0.11~mJy~beam$^{-1}$). 2.1 and 5.0~GHz combined image contours are 15, 20, 25, 30 and 33~$\sigma$ ($\sigma = $~0.06~mJy~beam$^{-1}$). 2.1, 5.0 and 9.0~GHz combined image contours are 3, 4, 5, 6, 7, 8, 9 and 10~$\sigma$ ($\sigma = $~0.05~mJy~beam$^{-1}$).}
\label{fig:contImages2016}
\end{figure*}

\begin{figure*}
\centering
\includegraphics[trim=0 0 0 0, width=.4\textwidth]{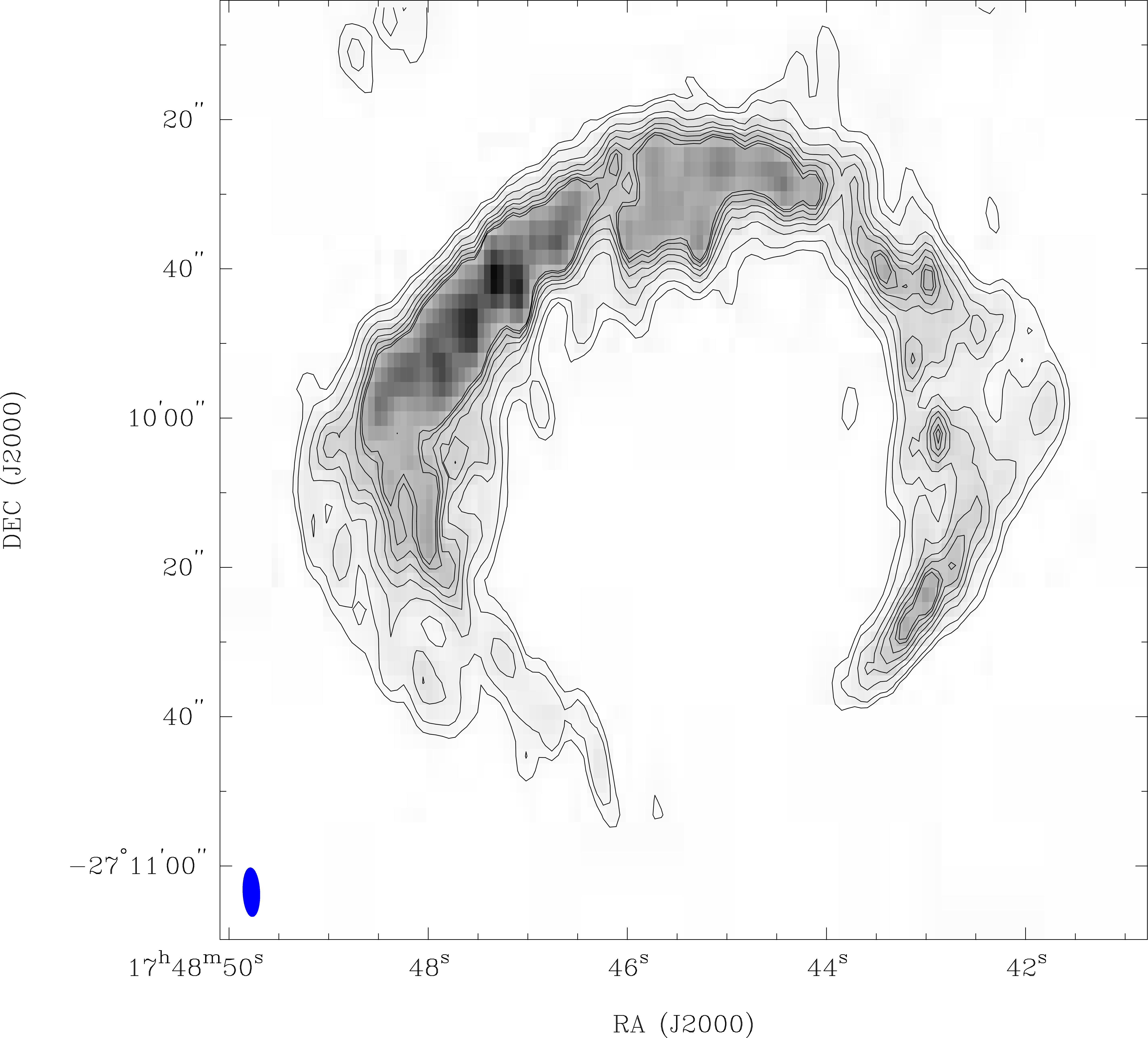}
\includegraphics[trim=0 0 0 0, width=.4\textwidth]{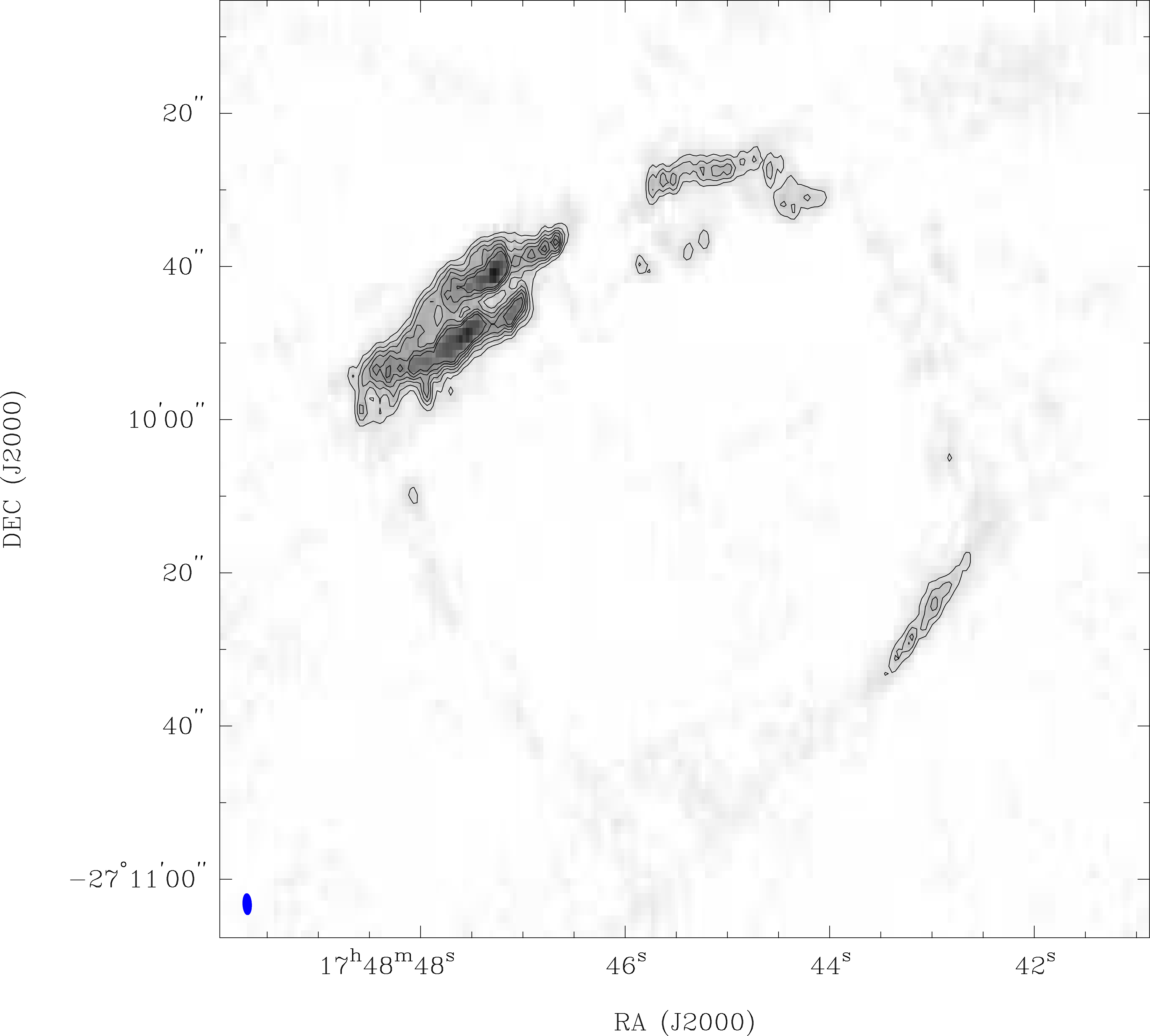}
\caption{Radio images produced from the 2017 observations of \g1\ using the \ac{ATCA}. The left image is 2.1~GHz, with the right being 5~GHz. The ellipse in the lower-left corner of each is representative of the beam size (Details in Table \ref{table:contImageDetails}). Contour levels for the 2.1~GHz image are 3, 5, 10, 15, 20, 25, 30 and 33~$\sigma$ ($\sigma = $~0.12~mJy~beam$^{-1}$). 5.0~GHz contours are 2, 3, 5, 6 and 7~$\sigma$ ($\sigma = $~0.13~mJy~beam$^{-1}$).}
\label{fig:contImages2017}
\end{figure*}

The \textsc{miriad} \citep{miriad} software package was used to reduce the data in multi-frequency synthesis mode, with the deconvolution being completed with the \textsc{invert}, \textsc{mfclean} and \textsc{restor} tasks, and primary beam correction with the \textsc{linmos} task with shadowed data flagged out. Figures~\ref{fig:contImages2016} and \ref{fig:contImages2017} show the final continuum images at each of observed frequencies (2.1, 5 and 9~GHz), as well as combined (2.1 and 5~GHz and 2.1, 5 and 9~GHz). The two combination images were synthesised using a restricted \textit{uv} range within the \textsc{invert} task to ensure appropriate (common) \textit{uv} coverage. \change{These images are the most sensitive (R.M.S Noise of 0.05/0.06~mJy~beam$^{-1}$) and prove to be ideal for our expansion study (Section~\ref{section:expansion}), and localised spectral index study (Section~\ref{section:specIndex}). However, we acknowledge that the physical meaning of such images is ambiguous, since different emission mechanisms are contributing emission on different scales at different frequencies.} 

The 2016 images at 2.1 and 5~GHz were created using a \textsc{robust} of --1, the 9~GHz image using a \textsc{robust} (`Briggs Weighting') of 0, and the two combination images with a \textsc{robust} of --1. All images were phase self-calibrated using the \textsc{selfcal} task. The 2017 images are all prepared using a \textsc{robust} of --1. Images in Figures~\ref{fig:contImages2016} and \ref{fig:contImages2017} were produced using the \textsc{cgdisp} task, and analysed using the \textsc{karma} software package \citep{kvis}.

The \HI\ absorption study in Section~\ref{section:absorption} was completed using the Stokes~I image cube produced by \textsc{miriad} with its \textsc{invert}, \textsc{clean}, \textsc{restor}, \textsc{uvmodel}, \textsc{uvlin} --- subtracting continuum using a linear model based on 1900 line-free channels --- tasks. 

The expansion study shown in Section \ref{section:expansion} was completed using the Stokes~I images produced by \textsc{miriad} with its \textsc{invert}, \textsc{mfclean}, \textsc{restor}, \textsc{linmos} tasks and shell profiles measured using the \textsc{cgslice} task. Figures~\ref{fig:expansionProfiles}, \ref{fig:expansionAll} and \ref{fig:expansionVectors} were created using the \textsc{matplotlib} \textsc{python} library \citep{matplotlib}. 

The polarisation study in Section \ref{section:polarisation} was completed using the Stokes I, Q and U images at 5~GHz produced by \textsc{miriad} with its \textsc{invert}, \textsc{mfclean}, \textsc{restor} and \textsc{impol} tasks. The images in Figure~\ref{fig:rm+pi} \change{were} produced by smoothing to a common resolution of 9\arcsec\ and \change{then} overlaying the polarisation vectors on a continuum image using the \textsc{cgdisp} task. The \ac{RM} in this section was measured by splitting the 2~GHz bandwidth into four equal bands. 

The spectral index and spectral index map in Section \ref{section:specIndex} were completed using the Stokes~I images produced by \textsc{miriad} with its \textsc{invert}, \textsc{mfclean}, \textsc{restor}, \textsc{linmos} and \textsc{mfspin} tasks. Figure~\ref{fig:SpecIndexMap} was created using the \textsc{cgdisp} task. The spectral index map is created using a combined 2.1, 5 and 9~GHz image, with the \textit{uv} coverage tapered to ensure that all frequencies cover the same \textit{uv} range. This resulted in better sensitivity over the wider frequency range at the expense of somewhat lower resolution.

To complement our ATCA data, we examined preliminary $^{12}$CO(1-0) and $^{13}$CO(1-0) data from the Mopra Southern Galactic Plane CO Survey \citep{Burton:2013}, taken between 2013 and 2018 by the 22-m Mopra radio telescope, located in the Warrambungles National Park, Australia. The full survey data release will cover longitudes of $-$110$^{\circ}<l<$11$^{\circ}$ and latitudes of $|b| < 1^{\circ}$, with extensions in selected regions of interest \citep{Braiding:2018}. The full survey also includes the C$^{18}$O(1-0) and C$^{17}$O(1-0) isotopologue transitions, however these were not available for this investigation. The specific \ac{CMZ} data presented in this paper are preliminary and will be publicly released by \citet{Blackwell:2018}, who also outline the full data reduction process. 

Mopra CO data have a 36\arcsec\ angular resolution. The Mopra spectrometer, MOPS, has eight 4096-channel dual-polarisation bands that deliver spectra with a velocity resolution of 0.1\,km~s$^{-1}$ when in `zoom'-mode. The full velocity-range of the CO data scrutinised in our analysis is $|v_{LSR}|<300$\,km~s$^{-1}$, encompassing all of the known molecular components within the \ac{CMZ}.

\begin{table*}
	\centering
	\caption{2016 and 2017 ATCA observation details of \g1. This table includes the date of the observations, the array configuration, the number of channels, the bandwidth in MHz and the frequency in MHz.}
	\label{table:obsDetails}
	\begin{tabular}{@{}lcccc@{}}
		\hline
Date &   Array & Channels & Bandwidth &  Frequency $\nu$   \\
 ~      &    Configuration    &    ~    &  (MHz)    &    (MHz) \\
		\hline
       26$^{th}$-27$^{th}$ January 2016 & EW352 & 5121 &  2.5  &  1421   \\
       26$^{th}$-27$^{th}$ January 2016 & EW352 & 2049 &  1    &  1610, 1664, 1666, 1719   \\ 
       26$^{th}$-27$^{th}$ January 2016 & EW352 & 2049 &  2048 &  2100, 5000, 9000    \\
       8$^{th}$-9$^{th}$ March 2016     & 6B    & 2049 &  2048 &  2100, 5000, 9000   \\
       20$^{th}$-21$^{st}$ May 2017     & 6A    & 5121 &  2.5  &  1421  \\
       20$^{th}$-21$^{st}$ May 2017     & 6A    & 2049 &  1    &  1610, 1664, 1666, 1719  \\
       20$^{th}$-21$^{st}$ May 2017     & 6A    & 2049 &  2048 &  2100, 5000, 9000   \\
		\hline
	\end{tabular}
\end{table*}

\begin{table*}
	\centering
	\caption{2016 and 2017 ATCA image details of \g1. Includes the year of the observations, the Frequency in GHz, the R.M.S. Noise in mJy~Beam$^{-1}$, the Synthesized Beam size and Position Angle, and the Robust Weighting scheme used during imaging (where lower values minimise sidelobes, and larger values optimise signal to thermal noise).}
	\label{table:contImageDetails}
	\begin{tabular}{@{}cccccc@{}}
		\hline
Year & Frequency $\nu$ & R.M.S. Noise & Synthesised & Position & Robust\\
~ &  (GHz) & (mJy Beam$^{-1}$)  & Beam & Angle & ~ \\
        \hline
       2016 & 2.1                   & 0.03 & 6.04\arcsec $\times$ 2.41\arcsec & --0.8\degr & --1  \\
       2016 & 5.0                   & 0.10 & 2.64\arcsec $\times$ 1.15\arcsec & --5.1\degr & --1 \\ 
       2016 & 9.0                   & 0.11 & 2.28\arcsec $\times$ 0.98\arcsec & --6.2\degr & 0  \\
       2016 & 2.1 $+$ 5.0           & 0.06 & 3.63\arcsec $\times$ 2.12\arcsec & --4.1\degr & --1  \\
       2016 & 2.0 $+$ 5.0 $+$ 9.0   & 0.05 & 3.49\arcsec $\times$ 2.11\arcsec & --5.1\degr & --1 \\
       2017 & 2.1                   & 0.12 & 5.35\arcsec $\times$ 2.06\arcsec & +2.4\degr  & --1 \\
       2017 & 5.0                   & 0.13 & 2.81\arcsec $\times$ 1.01\arcsec & +3.2\degr  & --1  \\
        \hline
	\end{tabular}
\end{table*}

\section {Results and Analysis} 

\subsection{Absorption} 
 \label{section:absorption}
\begin{figure*}
\centering
\includegraphics[width=\textwidth]{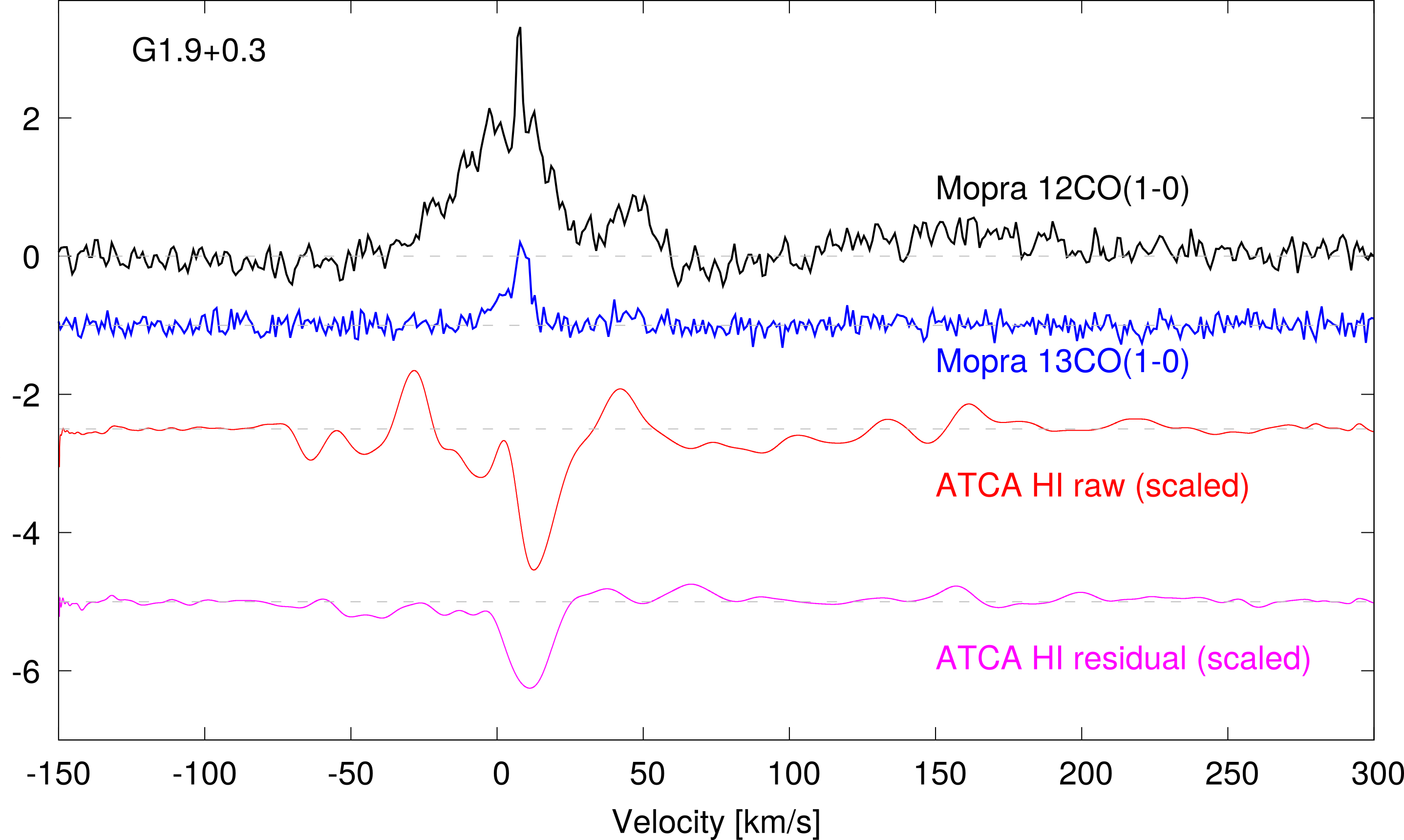}
\caption{Mopra $^{12}$CO(1-0), $^{13}$CO(1-0) \citep{Blackwell:2018}, and ATCA \HI\ spectra from the position of \g1. Both a raw \HI\ spectrum (with emission and absorption components) and residual \HI\ spectrum (i.e. targeting absorption components in the \g1 radio continuum only) are shown. \HI\ data were smoothed using a B\'{e}zier method.}
\label{fig:SpectraCO}
\end{figure*}

Figure\,\ref{fig:SpectraCO} displays spectra for Mopra $^{12}$CO(1-0), $^{13}$CO(1-0) and ATCA \HI\ towards \g1. Both a raw \HI\ spectrum and a residual \HI\ spectrum, where the \HI\ spectrum from a neighbouring region is subtracted, are shown. Absorption previously observed by \citet{2014IAUS..296..197R} at $\sim$10\,km~s$^{-1}$ is visible in our residual \HI\ spectrum.

\citet[][]{2014IAUS..296..197R} attributed this gas component to local and Sagittarius arm gas. $^{12}$CO(1-0) and $^{13}$CO(1-0) emission indicates that molecular gas is also present towards \g1 at a line-of-sight velocity $\sim$10\,km~s$^{-1}$ and it may be associated with an atomic component corresponding to the \HI-dip present in the residual \HI\ spectrum. However, since the central line velocities are offset by $\sim$3\,km~s$^{-1}$ ($\sim$7\,km~s$^{-1}$ for CO, $\sim$10\,km~s$^{-1}$ for \HI), we make no firm conclusion regarding an association. \change{Additionally, the complex nature of the \ac{GC} makes it difficult to disentangle the foreground and background sources}

\citet{2014IAUS..296..197R} also found \HI\ absorption components at $-$50\,km~s$^{-1}$ and $+$150\,km~s$^{-1}$. Our ATCA residual \HI\ spectrum does not clearly show either feature, so we do not make any new firm conclusions regarding the foreground/background nature of line of sight \HI\ components, and assume a distance of $\sim$8.5\,kpc in our analysis.

\begin{figure*}
\centering
\includegraphics[width=0.8\textwidth]{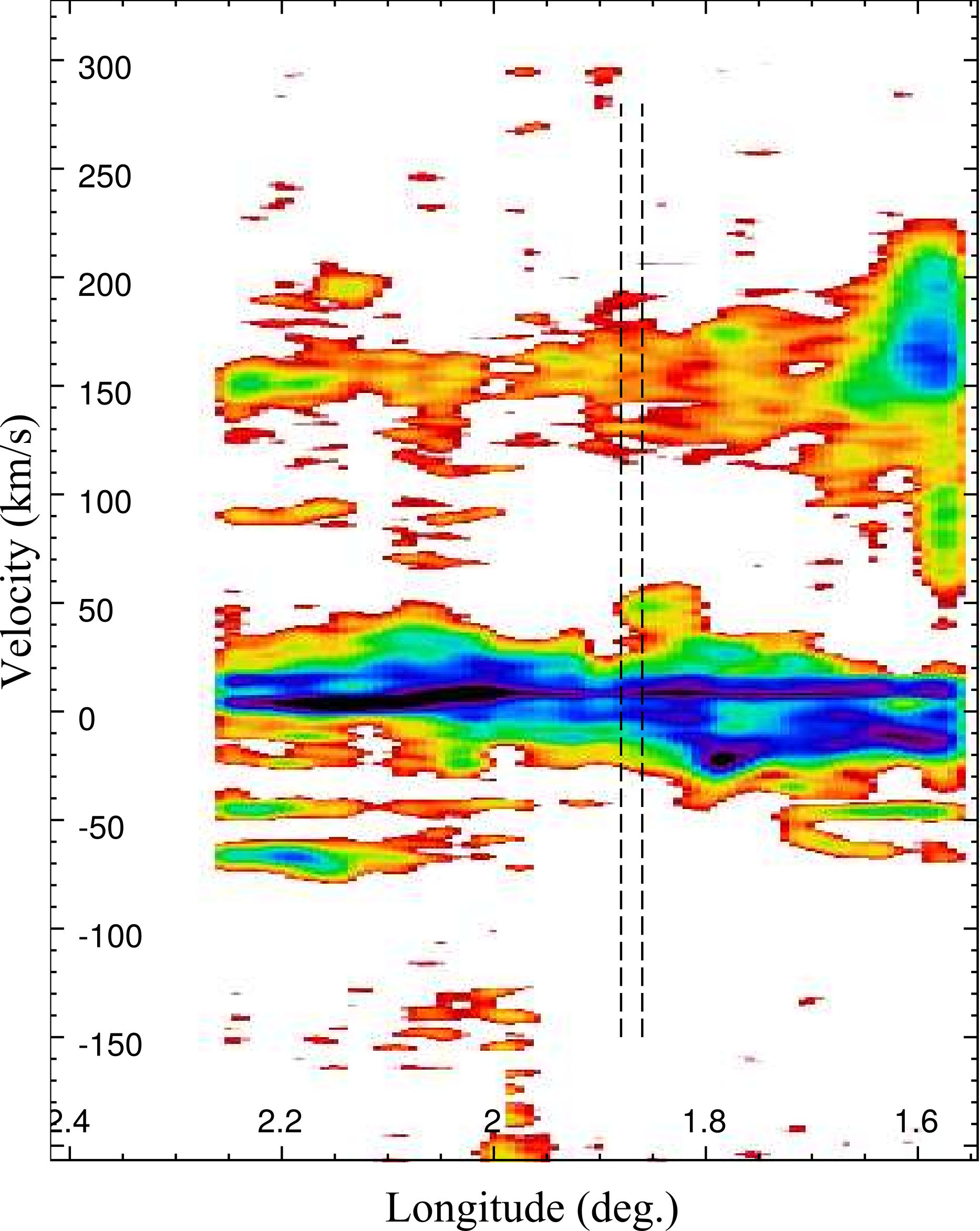}
\caption{Longitude-Velocity plot of $^{12}$CO(1-0) emission. Data are integrated between Galactic latitudes 0.31 to 0.34$^{\circ}$, corresponding to \g1. CO data are a preliminary result from the Mopra Central Molecular Zone CO Survey \citep{Blackwell:2018}. Vertical dashed lines indicate the longitudinal extent of \g1.}
\label{fig:COPVplot}
\end{figure*}

Figure\,\ref{fig:COPVplot} is a position-velocity image of Mopra CO(1-0) emission towards $\sim$0.9$^{\circ}$ of longitude encompassing \g1. This image shows near-zero and $\sim$50\,km~s$^{-1}$ components visible in the CO(1-0) spectrum (Figure\,\ref{fig:SpectraCO}), as well as gas at $+$150\,km~s$^{-1}$ that can not be discerned in Figure\,\ref{fig:SpectraCO}. As noted by \citet{2014IAUS..296..197R}, gas at $+$150\,km~s$^{-1}$ likely corresponds to the so-called `Feature-I', which is close to the inner Galactic centre and extends to be background to Sgr A*. As noted earlier, this feature is seen in \HI\ absorption by \citet{2014IAUS..296..197R}, but not confirmed in our analysis of ATCA \HI\ data. We further note that a tentative \HI\ emission component at $+$150\,km~s$^{-1}$ may exist in Figure\,\ref{fig:SpectraCO}. However, since `Feature-I' is very close to the Galactic Centre, a confirmation of this component would have little effect on the assumed \g1 distance of 8.5\,kpc.

\begin{table*}
	\centering
    \caption{Image details for the three images used in the \HI\ study.}
    \label{table:HIObservations}
    \begin{tabular}{@{}ccccc@{}}
    	\hline
 Image & R.M.S & 5$\sigma$ & Resolution & Synthesised\\
 type & (mJy beam$^{-1}$) & (mJy beam $^{-1}$) & (km s$^{-1}$) & Beam \& P.A. \\
        \hline
Continuum Map & 2.5 & 12.4 & N/A & 14.75\arcsec $\times$ 5.86\arcsec, 19.4\degr \\
Total Intensity Cube & 31.5 & 157.8 & 0.1 & 14.75\arcsec $\times$ 5.86\arcsec, 19.4\degr  \\
Spectrum Cube & 3.5 & 17.5 & 1 & 14.75\arcsec $\times$ 5.86\arcsec, 19.4\degr  \\
		\hline
   \end{tabular}
\end{table*}

\subsection{Expansion} \label{section:expansion}

Given the lack of known young Galactic \ac{SNR}s, confirmation of the age as well as the measurement of its expansion is very important for evolutionary studies. The calculation of \g1 age and expansion rate follows the method described in \citet{2014SerAJ.189...41D} and \citet{Roper:2018}

\begin{figure*}
\centering
\includegraphics[trim=0 0 0 0, width=0.75\textwidth]{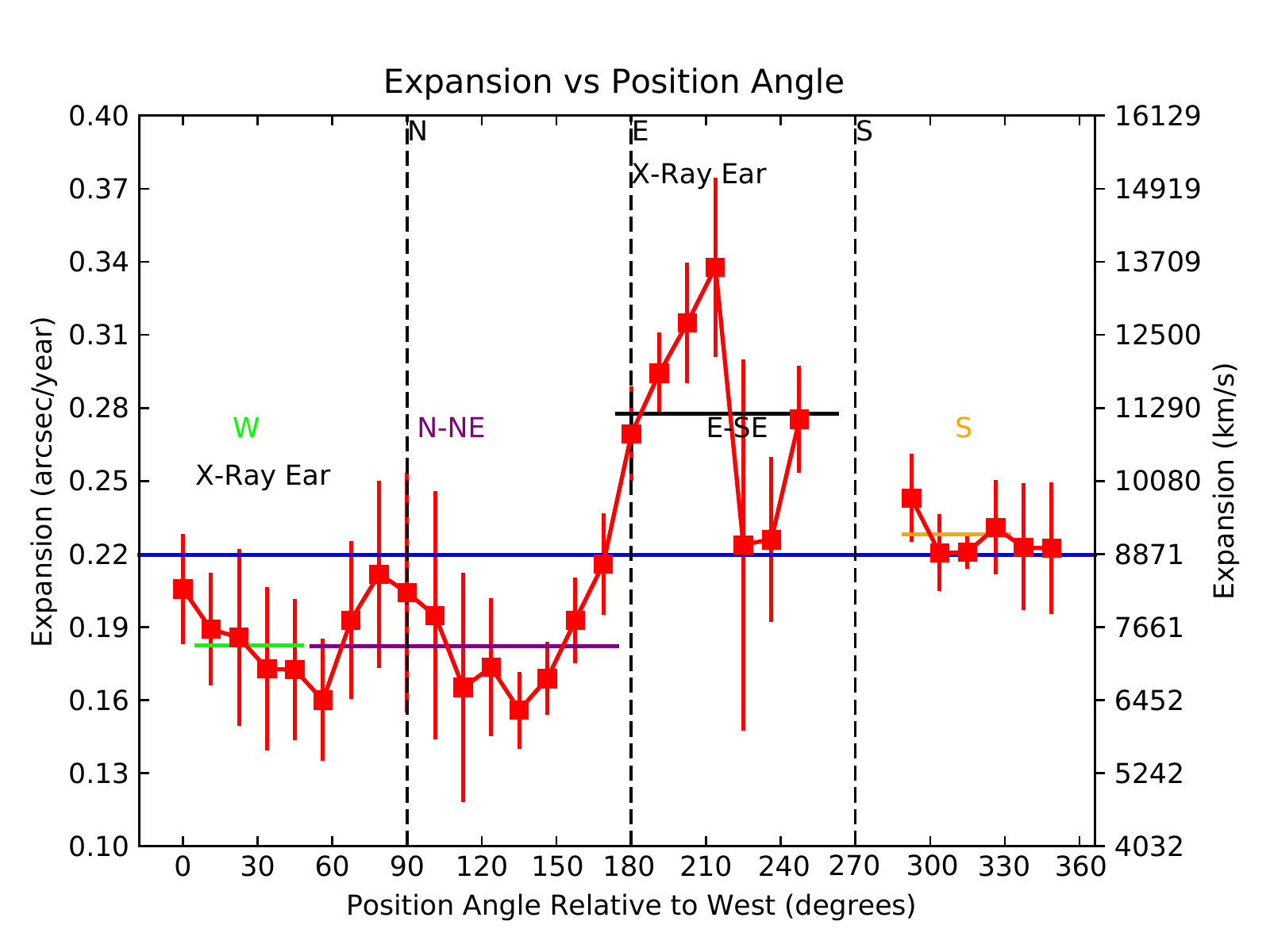}
\caption{Expansion rates measured using multiple images. We use all radio images accessible from 1984, 1985, 1987, 1989 and 2008 from the \ac{VLA} and 2009, 2016 and 2017 from the \ac{ATCA} (all synthesised to a beam size of 11.12\arcsec$\times$5.32\arcsec). The plot shows the position angle relative to West of which the expansion was measured along the X-axis, and the expansion measured along the Y-axis (measured in arcseconds per year on the left and kilometres per second on the right). }
\label{fig:expansionAll}
\end{figure*}

\begin{figure*}
\centering
\includegraphics[trim=0 0 0 0, width=0.75\textwidth]{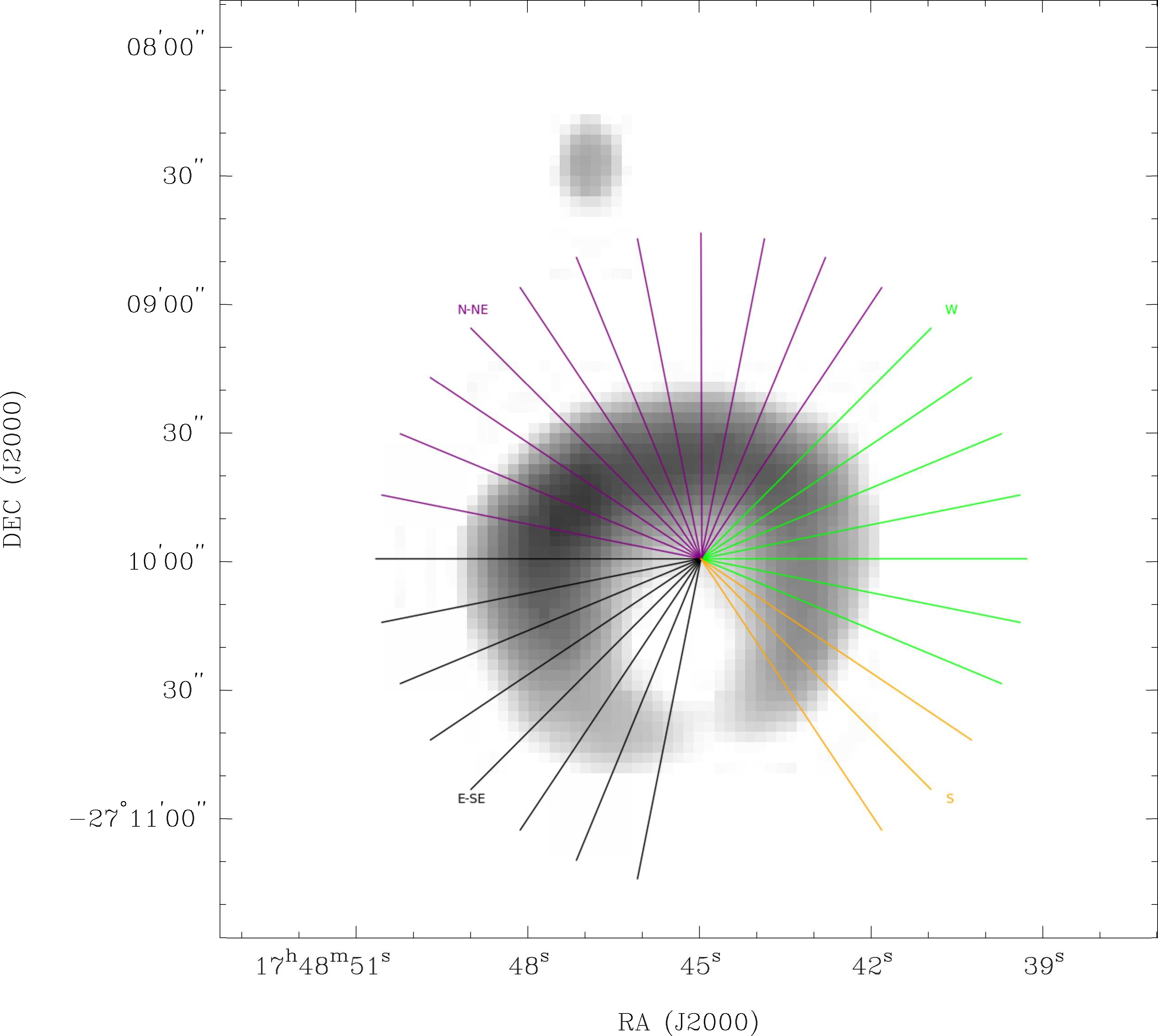}
\caption{Visualisation of the vectors used to calculate the expansion rates in Figure~\ref{fig:expansionAll}. The greyscale is the average of all epochs used. The coloured vectors correspond to the West / North-North-East / East-South East / South regions marked in Figure~\ref{fig:expansionAll}. Vectors extending towards the southern breakout region were discarded.}
\label{fig:expansionVectors}
\end{figure*}

Expansion is calculated across multiple radio-continuum images from between 1984 and 2017, produced from observations using both the \ac{VLA} and \ac{ATCA}. Details of the observations are in Table~\ref{table:expansionImages}. These images were smoothed/convolved to a single beam size (matched to the lowest resolution image -- 11.12\arcsec $\times$ 5.32\arcsec). They were then regridded to ensure that all images had the same center of RA(J2000)=17$^h$48$^m$45$^s$ and DEC(J2000)=--27\D10\arcmin6.7\arcsec and same pixel size. Shell profiles were then measured over 32 arcs, beginning at due west and continuing counter-clockwise (paralactic angle, see Figure~\ref{fig:expansionAll}). 32 arcs were chosen to avoid over-sampling the images, giving a shell profile every $\sim$11\D\ in \ac{SNR} shell azimuth, demonstrated in Figure~\ref{fig:expansionVectors}. 

Using the shell profiles measured from all 8 epochs over a single arc, we measure the distance from the \ac{SNR} centre to the peak radio brightness along the shell. These radii data points were plotted against corresponding years, and a least squares fit to the line was used to determine the expansion rate in arcseconds per unit of time. The residuals from the fitted line were used to establish the statistical uncertainty of this expansion rate. In this analysis, the southern break-out region where no clear shell profile could be discerned was excluded, demonstrated in Figure~\ref{fig:expansionVectors}.

\begin{figure*}
\centering
\includegraphics[trim=0 0 0 0, width=0.85\textwidth]{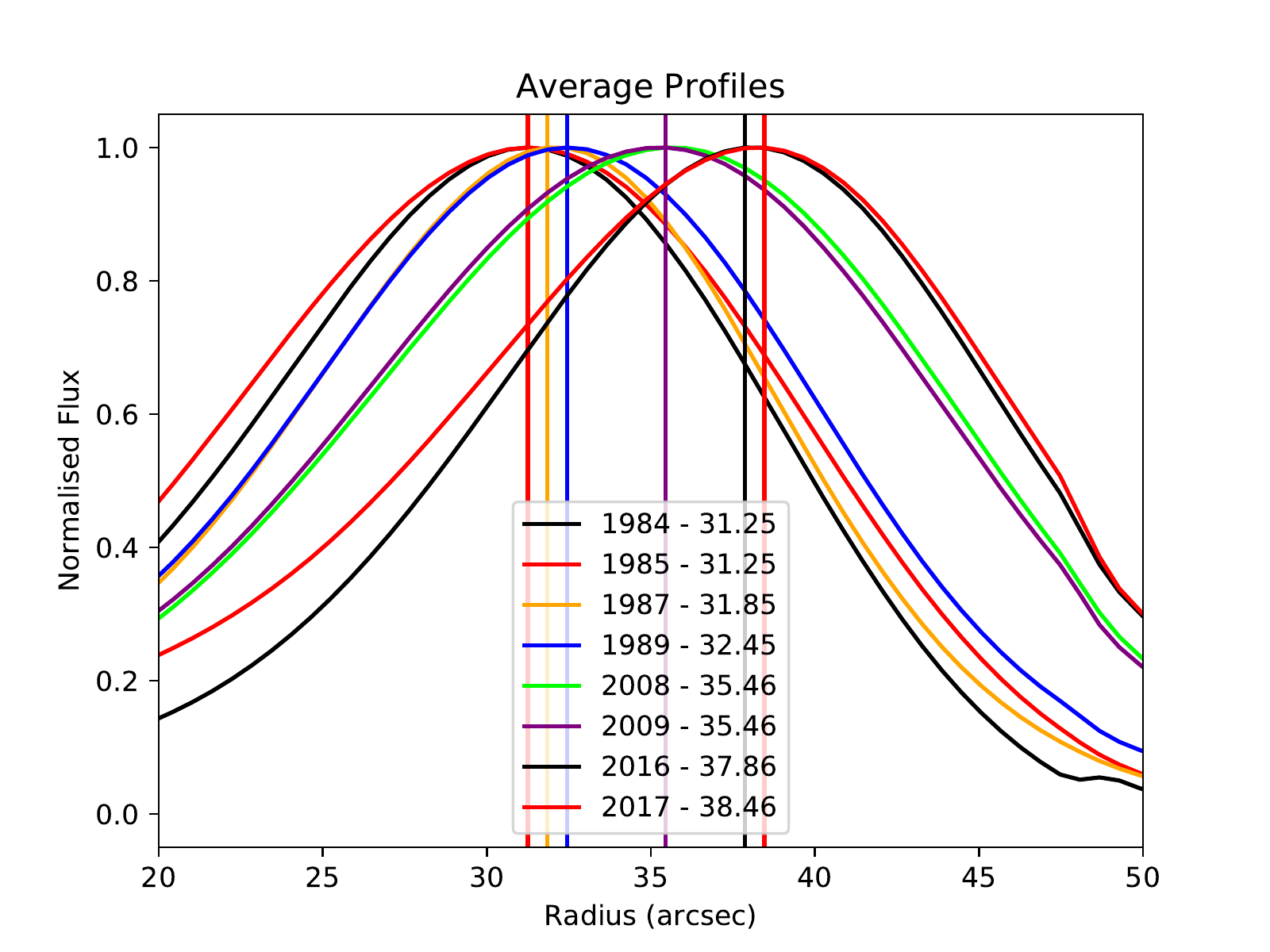}
\caption{Average shell profiles taken from images synthesised from observations taken from the \ac{VLA} in 1984, 1985, 1987, 1989 and 2008, and from the \ac{ATCA} in 2009, 2016 and 2017 (all synthesised to a beam size of 11.12\arcsec$\times$5.32\arcsec). Line profiles generally show the expected expansion, except in 2016-2017 where the lack of short spacings would seem to be producing a larger average shell profile than expected.}
\label{fig:expansionProfiles}
\end{figure*}

Once the entire \ac{SNR} (at each given observation date) had been measured and fitted lines produced, we estimated the expansion of \g1 (in arcseconds per year, percentage per year and kilometres per second), and finally, the approximate free-expansion age. The former is shown in Figures~\ref{fig:expansionProfiles} and \ref{fig:expansionAll}. The areas on Figure~\ref{fig:expansionAll} marked with a horizontal green, purple, black and orange line and labelled ``W'', ``N-NE'', ``E-SE'' and ``S'' correspond to the areas introduced by \citet{2017ApJ...837L...7B}, demonstrated in Figure~\ref{fig:expansionVectors}. This allows us to directly compare our radio continuum expansion study to the X-ray estimates. We have included in Table~\ref{table:expansionResults} the mean expansion rates and age, as well as the maximum expansion rates and age based thereupon. As expected, these are the regions with the largest expansion rate at both wavebands. 

Overall, we have found that the \ac{SNR} \g1 has expanded between 1984 and 2017 at an average rate of (0.78$\pm$0.09)~per~cent~per~yr or ($\sim$8\,900$\pm$1\,200)~km~s$^{-1}$ which implies a free-expansion age of (142$\pm$19) yrs, dating this \ac{SNR} explosion to mid-to-late 19$^{\mathrm{th}}$ century. This result agrees with previous studies estimating an expansion rate of (0.642$\pm$0.049)~per~cent per yr by \citet{2017ApJ...837L...7B,2011ApJ...737L..22C} and is slightly faster than the (0.563$\pm$0.078)~per~cent per yr measured by \citet{2014SerAJ.189...41D}. 

\begin{table*}
	\centering
	\caption{Epochs analysed for use in expansion study of \g1.}
	\label{table:expansionImages}
	\begin{tabular}{@{}lccccccc@{}}
		\hline
Observing & Project &  Telescope & Array Configuration & Bandwidth & Frequency   & Original Synthesised & Original Position\\
 Date     & Code    &    ~       &    Configuration    &  (MHz)    & $\nu$ (GHz) & Beam & Angle \\ 
		\hline
       26/05/1984 & AG0146 & \ac{VLA} & C & 50 & 4.8351 \& 4.8851 & 7.76\arcsec $\times$ 3.43\arcsec & --6.2\degr \\
       16/04/1985 & AG0184 & \ac{VLA} & B &  50 & 1.4649 \& 1.5149 & 2.78\arcsec $\times$ 1.11\arcsec & --5.5\degr\\ 
       22/02/1987 & AB0407 & \ac{VLA} & CD &  50 &  4.8351 \& 4.8851  & 10.05\arcsec $\times$ 9.27\arcsec & +64.1\degr \\
       23/06/1989 & AB0544 & \ac{VLA} & BC &  50 &  4.8351 \& 4.8851  & 8.03\arcsec $\times$ 3.35\arcsec & --26.3\degr \\
       12/03/2008 & AG0793 & \ac{VLA} & C &  50  &  4.8351 \& 4.8851  & 2.78\arcsec $\times$ 1.11\arcsec & --5.5\degr \\
       20/01/2009 & C1952 & \ac{ATCA} & EW352 + 6C &  128    &  4.5440 \& 5.1840  & 11.12\arcsec $\times$ 5.32\arcsec & --0.8\degr \\
       19/02/2016 & C1952 & \ac{ATCA} & EW352 + 6B &  4096 &  2.1000 \& 5.0000  & 7.74\arcsec $\times$  3.51\arcsec & --5.5\degr \\
       20/05/2017 & C1952 & \ac{ATCA} & 6A &  4096 &  2.1000 \& 5.0000  & 1.80\arcsec $\times$ 0.68\arcsec & +2.4\degr \\
		\hline
	\end{tabular}
\end{table*}

\begin{table*}
	\centering
    \caption{Expansion study results - matched resolution (11.12\arcsec$\times$5.32\arcsec) using images synthesised from observations from 1984, 1985, 1987, 1989, 2008, 2009, 2016, 2017. The regions are as defined in Figure~\ref{fig:expansionAll}. }
    \label{table:expansionResults}
    \begin{tabular}{@{}lccccccccc@{}}
    	\hline
    ~ & Average & Maximum & Average & Maximum & Average & Maximum & Average & Minimum \\    
 Region & Expansion & Expansion & Expansion & Expansion & Expansion & Expansion & Age & Age \\
 ~ & (arcsec year$^{-1}$) & (arcsec year$^{-1}$) & (\% year$^{-1}$) & (\% year$^{-1}$) & (km~s$^{-1}$) & (km~s$^{-1}$) & (years) & (years) \\
 \hline
Overall & 0.22 $\pm$ 0.03 & 0.34 $\pm$ 0.08 & 0.78 $\pm$ 0.09 & 1.20 $\pm$ 0.23 & 8854 $\pm$ 1195 & 13616 $\pm$ 3075 & 142 $\pm$ 19 & 93 $\pm$ 7 \\
West & 0.18 $\pm$ 0.03 & 0.19 $\pm$ 0.04 & 0.65 $\pm$ 0.09 & 0.67 $\pm$ 0.11 & 7364 $\pm$ 1248 & 7627 $\pm$ 1462 & 155 $\pm$ 18 & 149 $\pm$ 15 \\
North & 0.18 $\pm$ 0.03 & 0.21 $\pm$ 0.05 & 0.56 $\pm$ 0.10 & 0.65 $\pm$ 0.16 & 7338 $\pm$ 1290 & 8539 $\pm$ 2059 & 178 $\pm$ 18 & 153 $\pm$ 11 \\
East & 0.28 $\pm$ 0.03 & 0.34 $\pm$ 0.08 & 0.84 $\pm$ 0.10 & 1.02 $\pm$ 0.23 & 11181 $\pm$ 1324 & 13616 $\pm$ 3075 & 119 $\pm$ 17 & 98 $\pm$ 7 \\
South & 0.23 $\pm$ 0.01 & 0.24 $\pm$ 0.02 & 0.90 $\pm$ 0.04 & 0.96 $\pm$ 0.06 & 9196 $\pm$ 545 & 9798 $\pm$ 731 & 111 $\pm$ 41 & 104 $\pm$ 31 \\
	\hline
	\end{tabular}
\end{table*}

\subsection{Polarisation} \label{section:polarisation}

The polarisation of a \ac{SNR} can be an additional clue towards its age, with young \ac{SNR}'s typically exhibiting a radially orientated magnetic field \citep{2012SSRv..166..231R}. At the same time, polarisation is an indicator of the emission mechanism of the \ac{SNR}, with the presence of polarisation indicative of non-thermal emission from high energy electrons. As a shock evolves and sweeps-up an increasing mass, the shock is expected to decelerate, giving rise to Rayleigh--Taylor instabilities \citep{Gull,Chevalier}. As this occurs, the magnetic field lines become increasingly disordered and toroidal, which is apparent from the disordered polarisation vectors \citep{1973MNRAS.161...47G, 2004MNRAS.348L..19J}. This is largely borne out by similarly young Type~Ia \ac{SNR}'s \citep{1987AuJPh..40..771M}, as well as those in the \acl{LMC} \citep[\acs{LMC};][]{2014MNRAS.440.3220B}. 

The \ac{ATCA}, by default, records the Stokes Q and U parameters required to calculate the polarisation vectors. This ensures that as long as the secondary calibrator is observed regularly, and the data are correctly calibrated, then the resultant polarisation map will be reliable. 

\begin{figure*}
\centering
\includegraphics[trim={0cm 9cm 0cm 8.3cm}, clip, width=\textwidth]{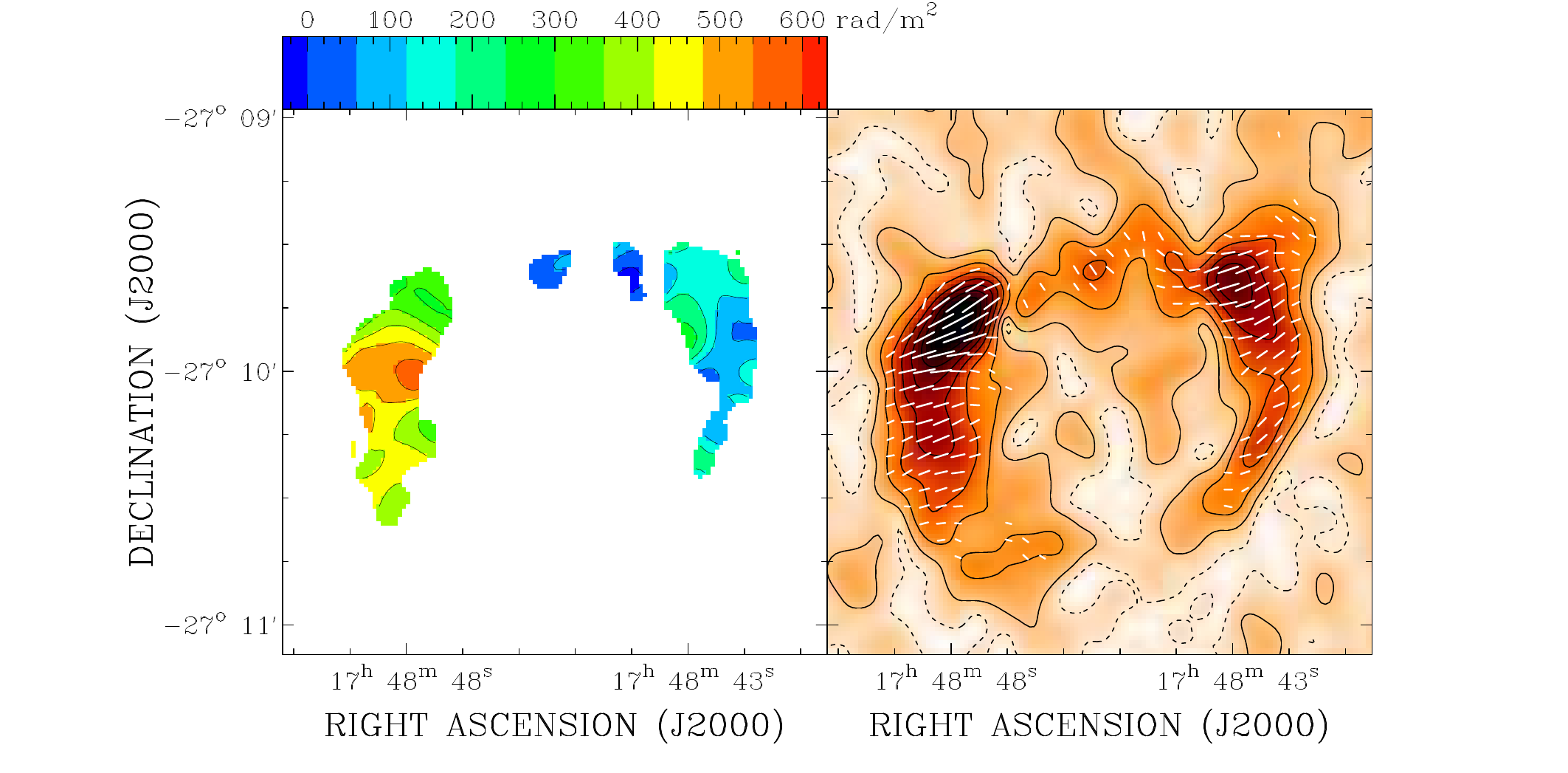}
\caption{\label{fig:rm+pi} Left: \ac{RM} map calculated for the \ac{SNR} \g1 after convolving the polarisation data to a common resolution of 9\arcsec. Displayed are only \acp{RM} calculated for a signal of more than $5\sigma$ in each band and for \acp{RM} with statistical errors of less than 20~rad~m$^{-2}$. Right: Polarized intensity map at 5~GHz integrated over the whole band. The vectors indicate the magnetic field at the point of origin projected to the plane of the sky. The vectors have been corrected for Faraday rotation.}
\end{figure*}

The full resolution polarisation images and the \ac{RM} map have very little signal-to-noise ratio. For a better analysis we therefore convolved the Q and U maps in the four bands to a common resolution of 9\arcsec. We re-calculated maps of polarised intensity (PI) and polarisation angle and the resulting \ac{RM} at the resolution of 9\arcsec. The resulting \ac{RM} map and an integrated polarised intensity map across the whole band around 5~GHz are displayed in Figure~\ref{fig:rm+pi}-left. We also corrected the observed polarisation angles for Faraday rotation and added the corrected B-vectors to the PI map in Figure~\ref{fig:rm+pi}-right.

In the \ac{RM} map in Figure~\ref{fig:rm+pi}-left we can see that the eastern (left) shell is dominated by high positive \acp{RM} of about +400 to +600~rad~m$^{-2}$, while the western (right) shell shows mostly low positive \acp{RM} of about 100 to 200~rad~m$^{-2}$. To quantify the distributions of \acp{RM} on both of the shells we plot RM values as a function of Right Ascension (RA) in Figure~\ref{fig:rmshell}. Dashed lines represent the mean values of the eastern and western shells, 411~rad~m$^{-2}$ and 123~rad~m$^{-2}$ respectively. The two concentrations of \acp{RM} at about $44^s$ and just below $46^s$ belong to the weakly polarised northern part of the shell (see left panel in Figure~\ref{fig:rm+pi}). In Figure~\ref{fig:rm+pi}-right the derived magnetic field vectors in the southern parts of the shells seem to be mostly parallel to the shock normal (parallel from now on), with an average angle of about $20^\circ$ to the RA axis. To the north, the eastern shell shows a departure from the parallel magnetic field at a Declination of about $-27^\circ 10'$ exhibiting close to tangential field (perpendicular to the shock normal; perpendicular from now on). This change in the field structure also coincides with a region of elevated RM value. The parallel magnetic field seems to continue further north for the Western shell, but then changes towards the weakly polarised blobs in the northern part of the shell. This difference in intrinsic magnetic field directions projected to the plane of the sky might be caused by possible interaction of the \ac{SNR} shell with molecular material in the north.

\begin{figure*}
\centering
\includegraphics[trim={0cm 5cm 0cm 5cm}, clip, width=\textwidth]{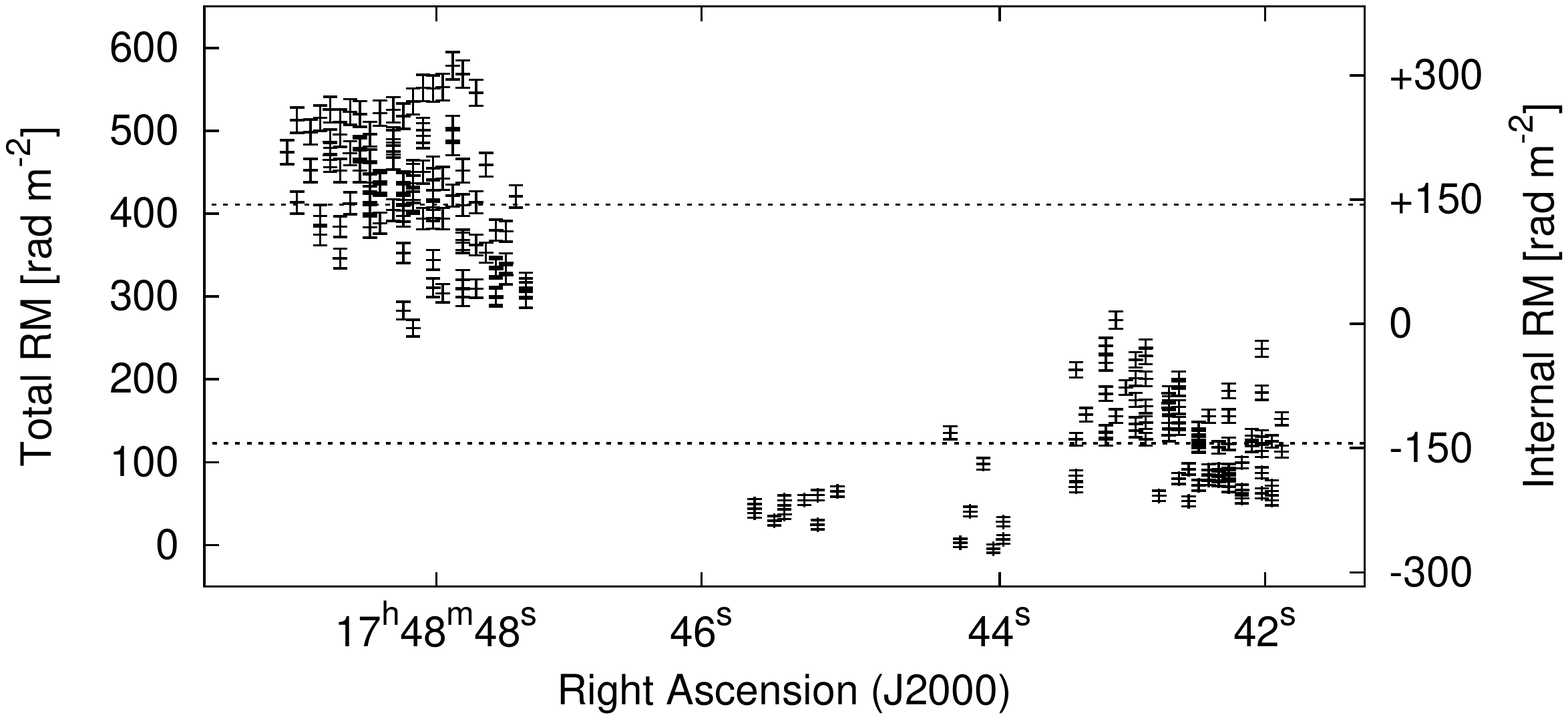}
\caption{\label{fig:rmshell} Plot of \acp{RM} observed for the \ac{SNR} \g1 as a function of Right Ascension. The dashed lines indicate the average \ac{RM} of 411~rad~m$^{-2}$ for the Eastern shell and 123~rad~m$^{-2}$ for the Western shell. The two concentrations of \acp{RM} at about 44$^s$ and just below 46$^s$ belong to the Northern shell, presumably interacting with molecular material and were taken into account. The right Y axis labels indicate the \acp{RM} after correcting for the assumed Galactic foreground \ac{RM} of 267~rad~m$^{-2}$.}
\label{fig:rmshell}
\end{figure*}

\subsection{Spectral Index} 
 \label{section:specIndex}

\begin{figure*}
\centering
\includegraphics[trim=0 0 0 0, width=.55\textwidth]{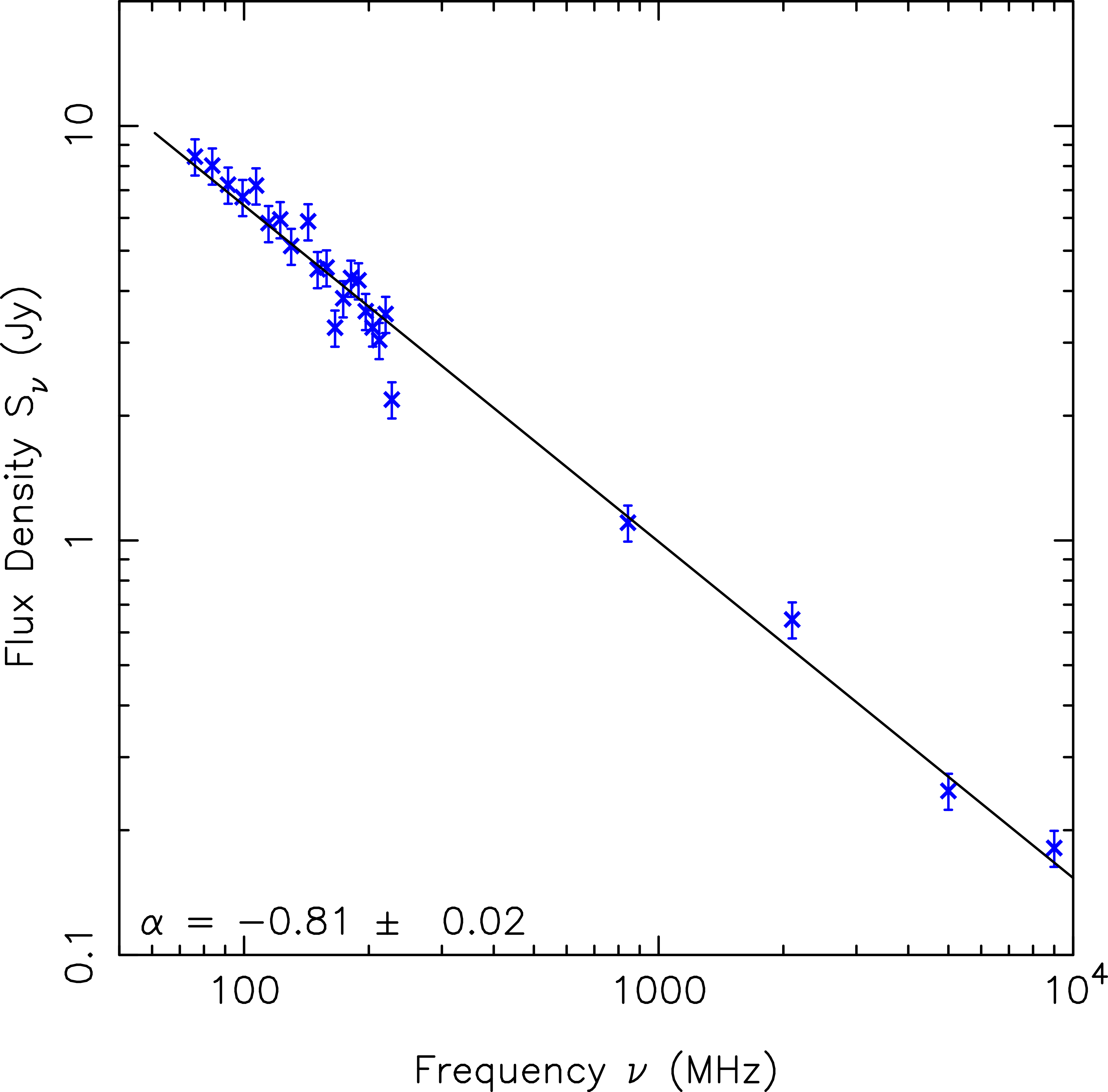}
\caption{Integrated spectral energy distribution of SNR G1.9+0.3 containing 20 flux densities from the \ac{GLEAM} survey (Hurley-Walker et al. submitted) (See Table~\ref{table:fluxDensities} for the individual flux density measurements used) from 76~MHz to 227~MHz, one 843~MHz flux density measured using the \ac{MOST} (with the flux density scaled based on the radio brightening found by \citet{2008MNRAS.389L..23M}) and the 2.1, 5.0 and 9.0~GHz flux densities measured using the 2016 \ac{ATCA} observations.} 
\label{fig:SpecIndex}
\end{figure*}

We estimated the spectral index in Figure~\ref{fig:SpecIndex} by fitting a line to the 20 flux densities, integrated over the source, between 76.155~MHz and 227.195~MHz measured by the \ac{GLEAM} project using the \acl{MWA} \citep[\acs{MWA};][]{2017MNRAS.464.1146H}, an 843~MHz flux density taken from \citet{2008MNRAS.389L..23M} measured with the \ac{MOST} and scaled by 12~per~cent to account for the brightening found within their paper, and the flux densities measured from our 2016 \ac{ATCA} observations. Our flux densities were measured from the 2016 \ac{ATCA} 2.1, 5 and 9~GHz images, where they had all been convolved to the same beam and pixel size. The 2.1~GHz image was then masked to 20$\sigma$ ($\sigma$ $=$ 0.16~mJy~beam$^{-1}$) and then used to mask the 5 and 9~GHz images to ensure the flux density measurements were taking into account the same pixels for all images. All flux densities used in Figure~\ref{fig:SpecIndex} are listed in Table~\ref{table:fluxDensities}. 

Using these flux densities, we obtain a spectral index of $-0.81~\pm$~0.02 --- shown in Figure~\ref{fig:SpecIndex} --- which is comparable to the spectral index of \mbox{--0.93~$\pm$~0.23} calculated by \citet{LaRosa:2000}. However, it is steeper than the spectral index calculated by \citet{2008MNRAS.387L..54G} of \mbox{--0.62$\pm$0.06.} \citet{2008MNRAS.387L..54G} note that their estimated spectral index differs from values in the literature, and instead suggest a spectral index of $\approx$--0.7, which is closer to the spectral index estimated here. With a spectral index of \mbox{--0.81~$\pm$~0.02}, \g1\ has one of the steepest spectral indexes known in our Galaxy and \acl{MCs} \citep[\acs{MCs}; \acused{MCs}][]{2017ApJS..230....2B,2019arXiv190811234M}. 

Such a steep spectral index is characteristic of young \acp{SNR} \citep{2014Ap&SS.354..541U}. The steep spectra of young SNRs can be caused by quasi-perpendicular magnetic field geometry \citep{2011MNRAS.418.1208B}, turbulent magnetic field amplification \citep{2019MNRAS.488.2466B}, Alfvenic drift effect \citep{2013MNRAS.433.1271J}, and by pure NLDSA effects which efficiently produce strong magnetic field amplification \citep{2017MNRAS.468.1616P}.

By using the radio spectrum from the \ac{ATCA}, \ac{MOST} and \ac{MWA} in conjunction with the distance of $\sim$~8.5~kpc and a diameter of 95\arcsec, we can calculate the total flux density at 1~GHz (0.99 Jy) surface Brightness ($0.60 \times 10^{-19}$ W~(m$^2$~Hz~SR)$^{-1}$) and luminosity between 10~MHz and 100~GHz (0.89$\times$10$^{26}$~W~Hz$^{-1}$).

\begin{figure*}
\centering
\includegraphics[trim=0 0 0 0, width=0.55\textwidth]{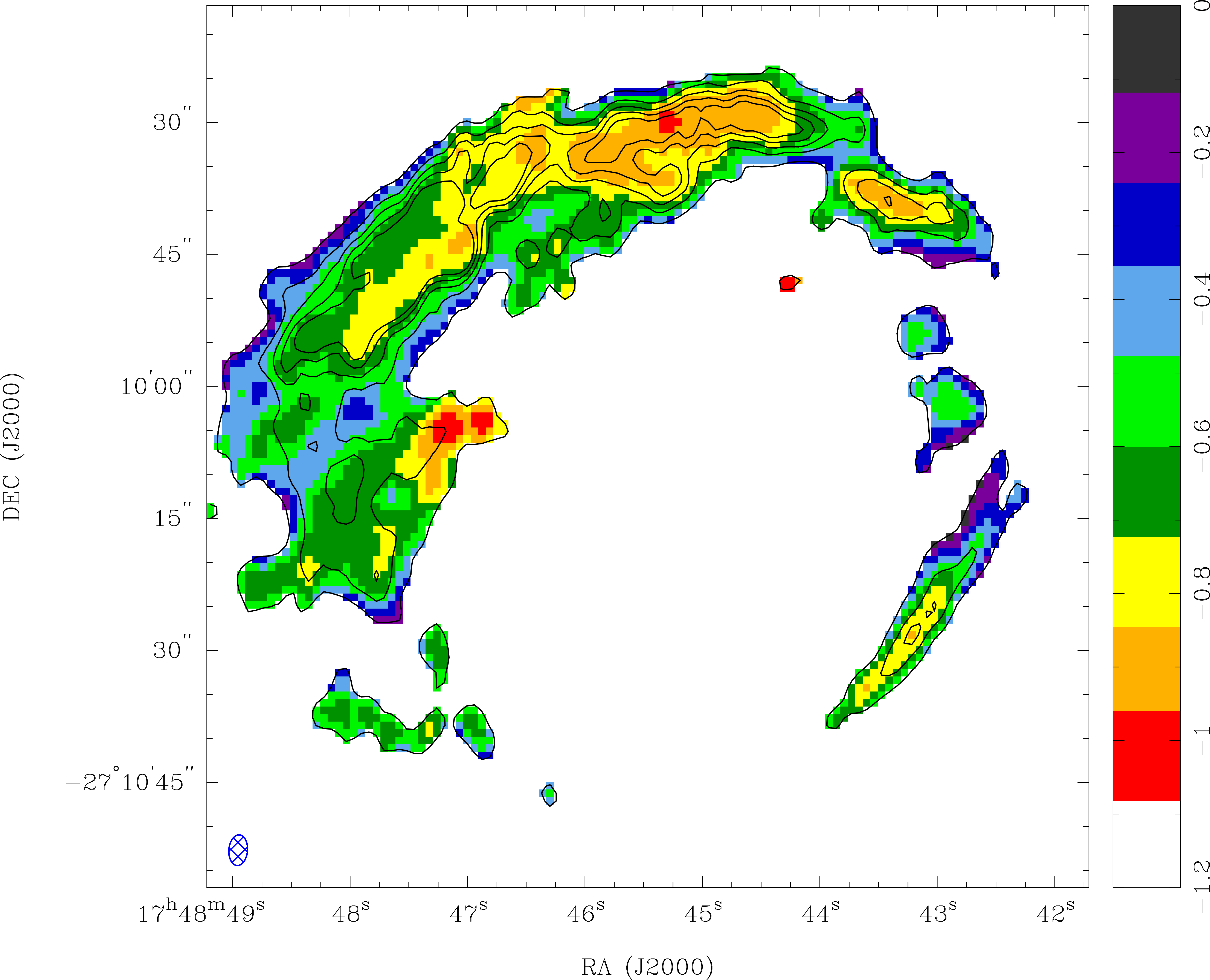}
\caption{Spectral index map produced from observations from 2016 using the \ac{ATCA}. This plot shows an average of --0.61, minimum value of --1.07 and standard deviation of 0.18. }
\label{fig:SpecIndexMap}
\end{figure*}

The spectral index map presented in Figure~\ref{fig:SpecIndexMap} is the product of the \textsc{miriad} task \textsc{mfspin}, with our 2.1, 5.0 and 9.0~GHz data combined into a single \textit{uv} file before imaging, with the \textit{uv} range tapered to ensure matching coverage across all bands. This image was then masked to 20$\sigma$ -- where $\sigma$ is measured to be 0.05~mJy~beam$^{-1}$, before the \textsc{miriad} task \textsc{mfspin} was run, allowing for the final 20$\sigma$ spectral index map in Figure~\ref{fig:SpecIndexMap} to be created. The steepest section of this spectral index map is measured at --1.07 in the northern region, where we also see somewhat randomised polarisation vectors (Figure~\ref{fig:rm+pi}) and extreme radio brightening in Section~\ref{section:brightening}. The average spectral index measured across the \ac{SNR} is \mbox{--0.61}, with a standard deviation of 0.18. 

We would not necessarily expect the spectral index presented in Figure~\ref{fig:SpecIndex} to be equivalent to the average spectral index measured from the spectral index map presented in Figure~\ref{fig:SpecIndexMap}\change{, as the spectral index presented in Figure~\ref{fig:SpecIndex} is effectively taking the average across the map unweighted by flux density, whereas the spectral index map in Figure~\ref{fig:SpecIndexMap} is effectively weighted by the flux density}.

\begin{table}
	\centering
    \caption{Flux densities, integrated over \g1\, measured using the \ac{MWA}, \ac{MOST} and \ac{ATCA}. All flux density measurements are assumed to have an error of 10~per~cent, with the \ac{MOST} 843~MHz flux density scaled up by 12~per~cent to account for the brightening found by \citet{2008MNRAS.389L..23M}}
    \label{table:fluxDensities}
    \begin{tabular}{@{}lcc@{}}
    	\hline
    Instrument & $\nu$ (MHz) & $S$ (Jy)  \\    
    \hline
    \ac{MWA} & 76.155 & 8.438 \\
    \ac{MWA} & 83.835 & 8.022 \\
    \ac{MWA} & 91.515 & 7.216 \\
    \ac{MWA} & 99.195 & 6.734 \\
    \ac{MWA} & 106.875 & 7.184 \\
    \ac{MWA} & 114.555 & 5.827 \\
    \ac{MWA} & 122.235 & 5.953 \\
    \ac{MWA} & 129.915 & 5.135 \\
    \ac{MWA} & 142.715 & 5.885 \\
    \ac{MWA} & 150.395 & 4.512 \\
    \ac{MWA} & 158.075 & 4.556 \\
    \ac{MWA} & 165.755 & 3.261 \\
    \ac{MWA} & 173.435 & 3.838 \\
    \ac{MWA} & 181.115 & 4.301 \\
    \ac{MWA} & 188.795 & 4.240 \\
    \ac{MWA} & 196.475 & 3.575 \\
    \ac{MWA} & 204.155 & 3.263 \\
    \ac{MWA} & 211.835 & 3.044 \\
    \ac{MWA} & 219.515 & 3.518 \\
    \ac{MWA} & 227.195 & 2.188 \\
    \ac{MOST} & 843.000 & 1.104 \\
    \ac{ATCA} & 2100.000 & 0.645 \\
    \ac{ATCA} & 5000.000 & 0.249 \\
    \ac{ATCA} & 9000.000 & 0.181 \\
	\hline
	\end{tabular}
\end{table}

\subsection{Radio Brightening}
\label{section:brightening}

\begin{figure*}
\centering
\includegraphics[trim={0 0 0 0}, width=0.55\textwidth]{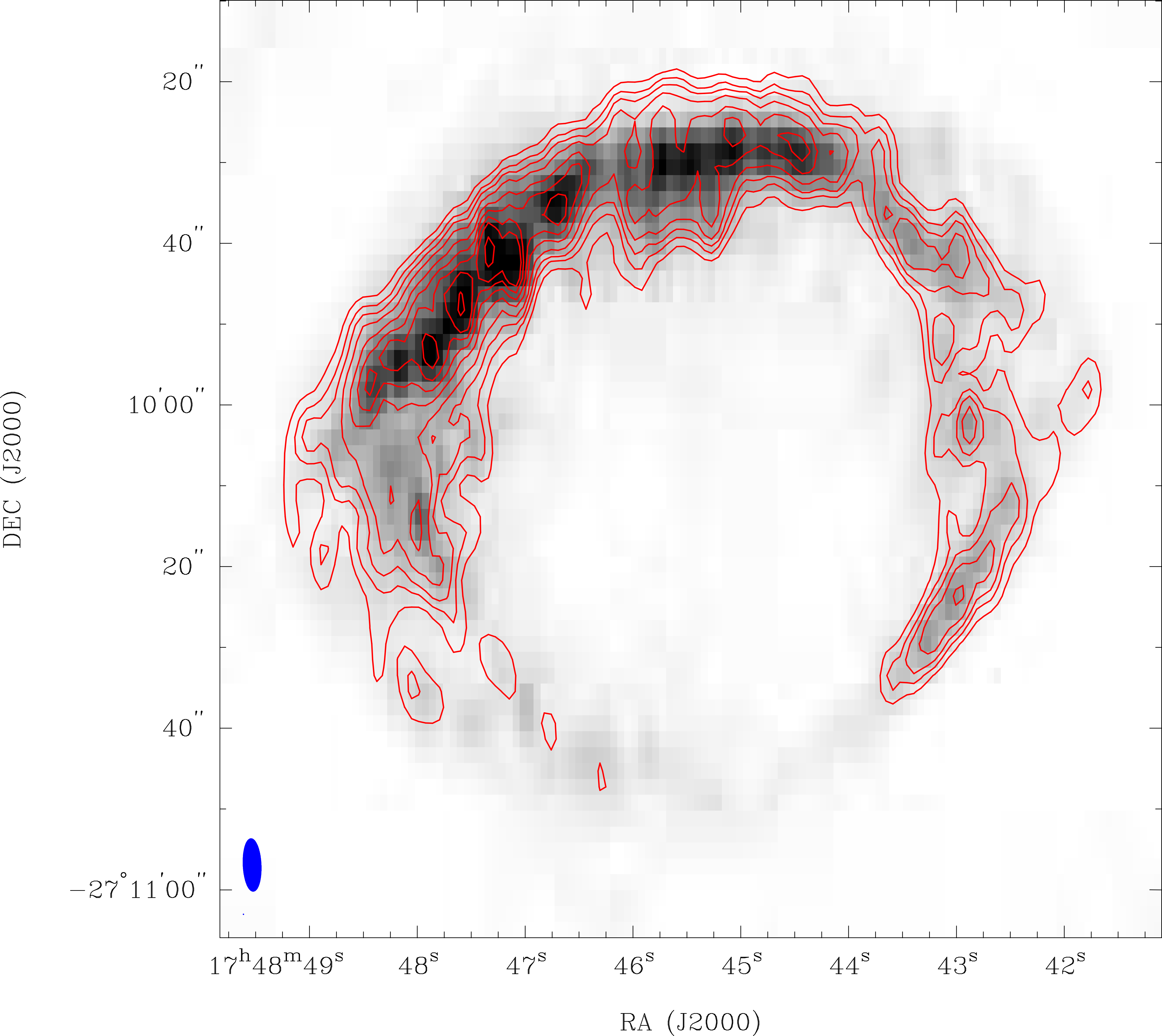}
\caption{2016 6B array observations at 2.1~GHz as the greyscale image, with contours taken from our 2017 6A observations at 2.1~GHz. Contour levels are 3, 5, 7, 10, 15, 20, 25, 30 and 40 $\sigma$ ($\sigma = 0.16$~mJy\,Beam$^{-1}$).}
\label{fig:expansion1617}
\end{figure*}

\begin{figure*}
\centering
\includegraphics[trim={0 0 0 0}, clip, width=0.55\textwidth]{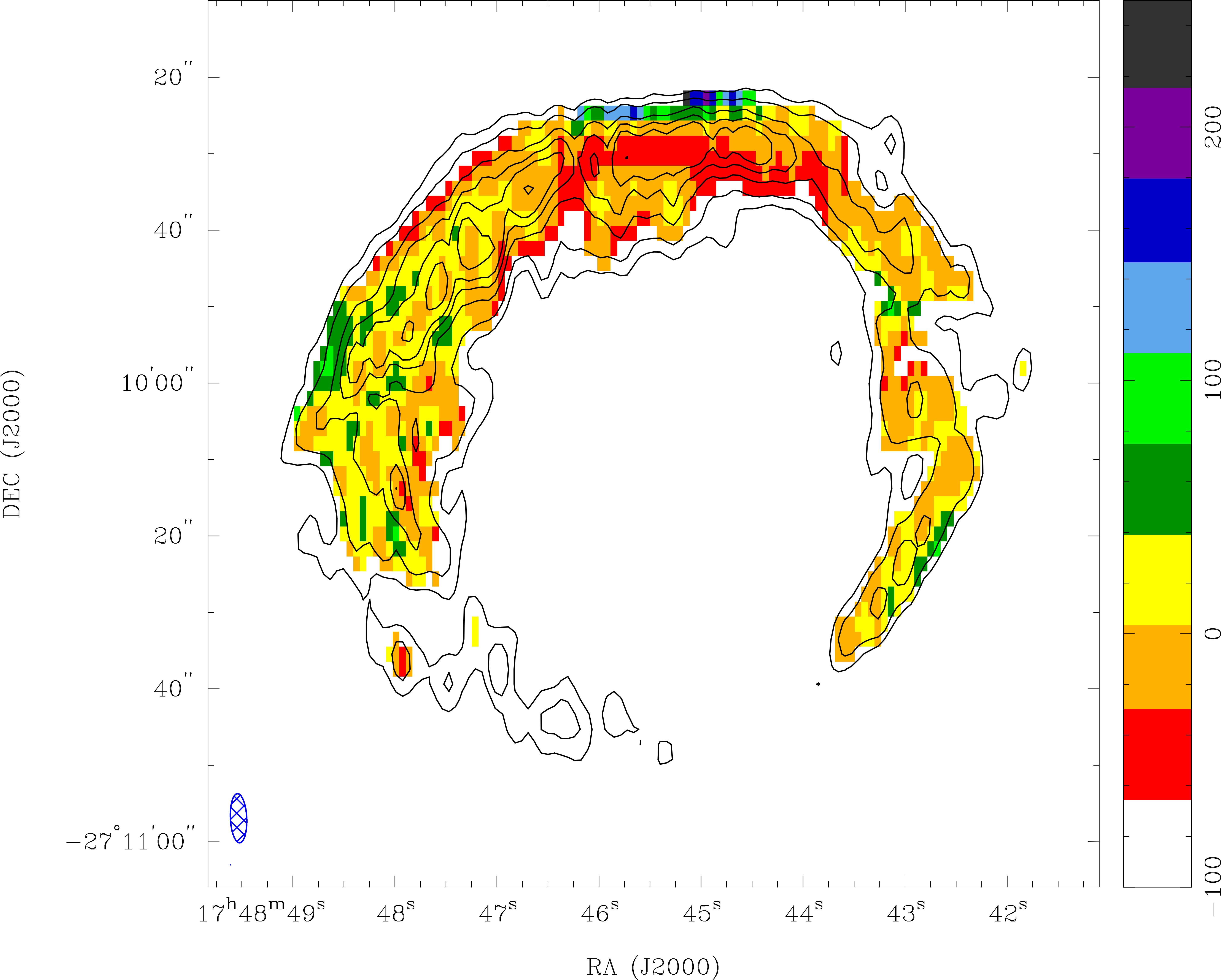}
\includegraphics[trim={0 0 0 0}, clip, width=0.75\textwidth]{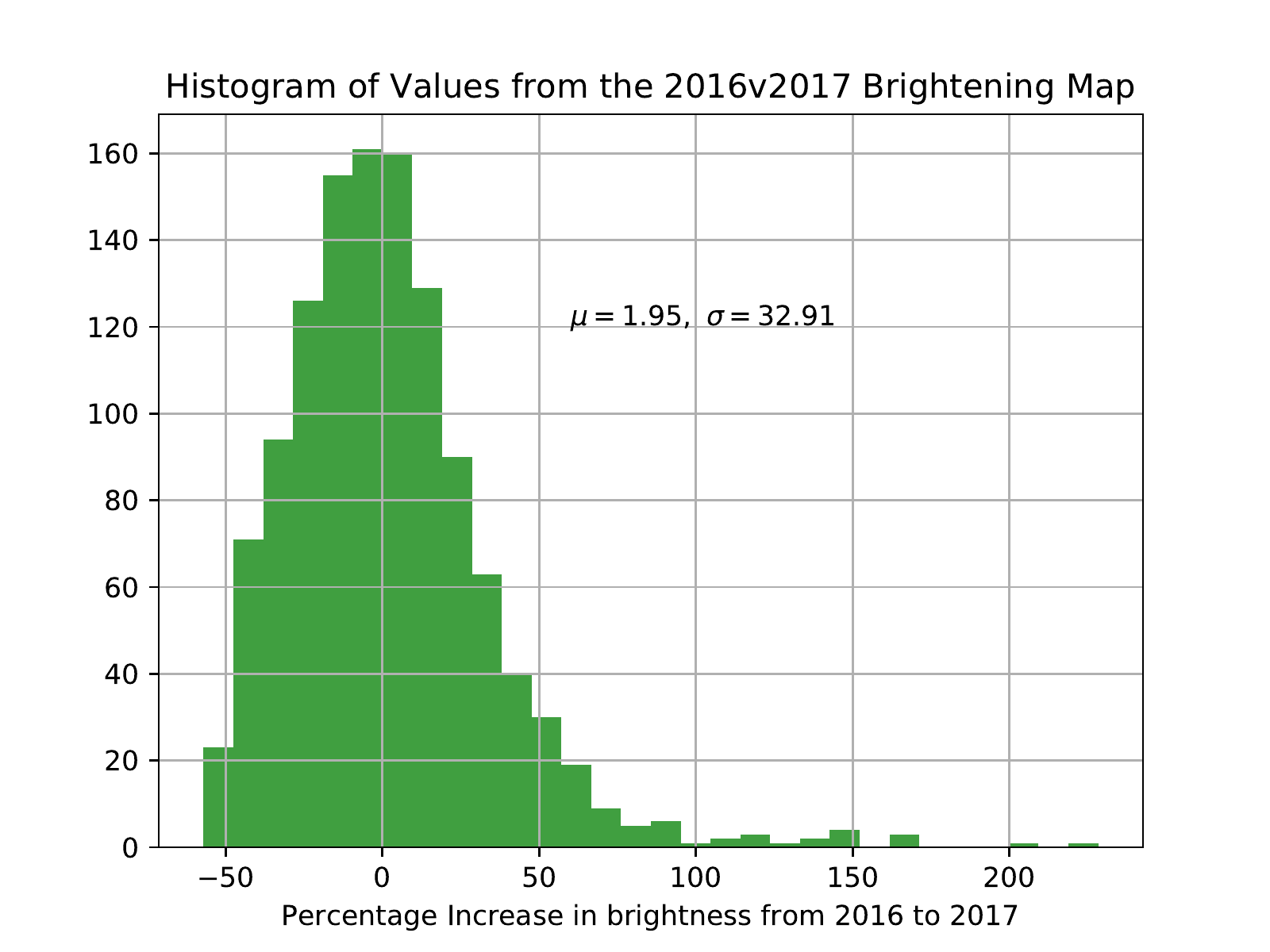}
\caption{Radio brightening map (top) produced from the 2016 6B array observations and 2017 6A array observations using the \ac{ATCA}. Both 2016 and 2017 observations were masked below 10$\sigma$ (2016 $\sigma =$0.16~$\mu$Jy~beam$^{-1}$, 2017 $\sigma=$0.28~$\mu$Jy~beam$^{-1}$). Average brightening in radio is found to be 1.95~per~cent with a minimum of --57.08~per~cent, maximum of 228.3~per~cent and a standard deviation of 32.31. Contours taken from the 2016 2.1~GHz 6B image at 5, 10, 20, 30, 50 and 80 $\sigma$ ($\sigma = 0.16$~mJy\,Beam$^{-1}$). Histogram (bottom) showing the counts of pixel intensities from the Radio Brightening Map (top). The region of highest flux density in the northern region shows a dimming due to the shift of the peak in the northern region moving further north. 
}
\label{fig:RadioBrightening}
\end{figure*}

In Figure\,\ref{fig:expansion1617}, a striking difference in the northern shock position was observed, with the 2017 \g1 image extending farther north than 2016 image. By taking our 2.1~GHz observations from March 2016 with the \ac{ATCA} in a 6B configuration and our 2.1~GHz observations from May 2017 with the \ac{ATCA} in a 6A configuration imaged to the same beam size, we can construct an image comparing the flux density levels from each. Figure~\ref{fig:RadioBrightening} is an image that shows the relative radio flux density brightening, calculated using Equation~\ref{eqn:fluxBrightening} across the 14 month period.

\begin{equation}
    \mathrm{Brightening} = \frac{S_{2017} - S_{2016}}{S_{2016}}\times100
    \label{eqn:fluxBrightening}
\end{equation}

We mask the resulting image so that any pixel is masked if it is below 10$\sigma$ in either the 2016 image ($\sigma =$0.16~mJy~beam$^{-1}$, or in the 2017 image  ($\sigma=$0.28~mJy~beam$^{-1}$).

The brightness percentage map in Figure~\ref{fig:RadioBrightening} has an average increase of 1.95~per~cent. There is a minimum of --57.08~per~cent, and a maximum brightening of 228.30~per~cent along the northern region where we found decreased expansion in Section~\ref{section:expansion}, a less-ordered polarisation field \textbf{as can be seen in Figure~\ref{fig:rm+pi}-Right} and a steep spectral index in Section~\ref{section:specIndex}.

By interpolating this 1.95~per~cent brightening rate from the measured 14 months to 12 months, we find an average brightening of (1.67$\pm$0.35)~per~cent~per~yr (consistent with the value of 1.8~per~cent predicted by simulations in \citet{2017MNRAS.468.1616P}), minimum brightening rate of $\sim$--50~per~cent~per~yr and a maximum brightening rate of $\sim$200~per~cent~per~yr. Our new average result is in line with the (1.22$^{+0.24}_{-0.16}$)~per~cent~per~yr measured by \citet{2008MNRAS.389L..23M} with their study over 20 years using the \ac{MOST}, and the $\approx$2~per~cent increase per year measured by \citet{2008MNRAS.387L..54G} with the \ac{VLA}. 

\section{Discussion}
\label{section:discussion}

\ac{SNR} \g1\ is unique, in that it is the youngest known Galactic Type~Ia \ac{SNR}. As such, it provides a window into the evolution of the young \acp{SNR}. While some of the observations demonstrated are within the expected and currently measured bounds -- Radio expansion rate and spectral index -- some observations differ. 

The \ac{RM} map shown in Figure~\ref{fig:rmshell} is indicative of one of the divergences from current observations. A similar behaviour for the distribution of \acp{RM} on the two opposite shells of an \ac{SNR} were found for the \ac{SNR} G296.5+10.0 by \citet{2010ApJ...712.1157H}. They found relatively constant \acp{RM} of opposite sign on both shells of the \ac{SNR}. As \g1 is estimated to be further away, and closer to the centre of our Galaxy, we would expect quite significant foreground \ac{RM}. Assuming a positive foreground \ac{RM} of about +267~rad~m$^{-2}$, the internal eastern and western shell \acp{RM} would be roughly matched with opposite signs of about $\pm144$~rad~m$^{-2}$ (see Figure~\ref{fig:rmshell}). We tried to analyse the foreground \ac{RM} by comparing \acp{RM} of pulsars and extra-galactic sources, at short angular distances from the \ac{SNR}. For the pulsars we are using the ``The Australia Telescope National Facility Pulsar Catalogue''  \citep{2005AJ....129.1993M}\footnote{\url{http://www.atnf.csiro.au/research/pulsar/psrcat/}}. We find two pulsars with known \ac{RM} of +421 and +916~rad~m$^{-2}$ within $1^\circ$ of \g1 \citep{2018ApJS..234...11H} with \ac{DM} distances of 5.2 and 4.2~kpc, respectively \citep{2017ApJ...835...29Y}. There are two linearly polarised background extra-galactic sources with \acp{RM} of +806 and +638~rad~m$^{-2}$ within $1^\circ$ of \g1 taken from a catalog of Faraday \ac{RM} of point sources published by \citet{2014RAA....14..942X}\footnote{\url{http://zmtt.bao.ac.cn/RM/searchGRM.html}}. These values indicate that the \acp{RM} in this direction are varying a lot and are in general positive with a high amplitude. A foreground \ac{RM} of +267~rad~m$^{-2}$ is certainly not unexpected, but is low compared to other \acp{RM} from the same direction. However, it also cannot be excluded that the difference in the \ac{RM} between the eastern and western shell is simply due to variations of the foreground \ac{RM}.

\citet{2010ApJ...712.1157H} showed that such a behaviour of the \ac{RM} in G296.5+10.0 can be explained by a \ac{SNR} expanding inside an azimuthal/toroidal magnetic field in the stellar wind of the progenitor star of the \ac{SN} explosion. In this case, magnetic field lines would be wrapped around the expanding \ac{SNR} in the equatorial region which in the projected picture will show up as a quasi-radial field inside the shell with the decrease of the in-plane component towards the edge of the remnant. This roughly corresponds to what we see in \g1 (see Figure~\ref{fig:rm+pi}-left) if the equatorial plane is tilted by $20^{\circ}$ with respect to the RA axis. From such a magnetic field configuration we would expect the \acp{RM} to have the same amplitude on both shells but opposite sign. \citet{2010ApJ...712.1157H} simulated the expansion of a late time stellar wind of the progenitor star in order to calculate the \ac{RM} that would be imposed on background linearly polarised emission. They studied several different progenitor stars and showed that the \ac{RSG} wind can be responsible for the levels of the \ac{RM} in G296.5+10.0. Using the same formalism, it is easy to demonstrate that this could also be a plausible scenario to explain the \ac{RM} distribution in \g1.

However, the \ac{RSG} wind and toroidal magnetic field hypothesis does not explain all aspects of this \ac{SNR}. 

First, to reach the size of \g1\ in this scenario, the age of the \ac{SNR} would need to be closer to $\sim$400 years old \citep{2005ApJ...630..892D, 2013A&A...552A.102T}, because of the expected shock speed of a \ac{SNR} expanding inside the stellar wind of the progenitor star. A super-luminous \ac{SN} explosion would be required to reach the expansion velocity and size observed in \g1. Additionally, the lower shock speed anticipated from this scenario would result in a lower maximum electron energy, which would be difficult to reconcile with the observed X-ray emission. The \ac{RSG}-scenario should produce more thermal X-ray emission than reported by \cite{2013ApJ...771L...9B}.

Second, the \ac{RSG}-scenario would make \g1 a twin of Cassiopeia~A, which would raise the question -- why are the respective morphologies so different?

Third, a toroidal magnetic field in the stellar wind of the progenitor star would be parallel to the shock, which makes particle acceleration at the shock very inefficient \citep{2014ApJ...783...91C, 2003A&A...409..563V} which in turn results in a lack of magnetic field amplification by the \acp{CR}. This contradicts the explanation of the synchrotron emission from the \ac{SNR}. Theoretical modeling requires a high magnetic field downstream of the forward shock with estimates ranging from 180~$\mu$G \citep{2019arXiv190602725B} to $>300~\mu$G \citep{2017MNRAS.468.1616P, 2018ApJ...855...59U} which in turn are in agreement with an equipartition magnetic field of $180-273~\mu$G \citep{2014SerAJ.189...41D}. 

Finally, simulations presented in \citet{2019arXiv190602725B} strongly suggest that radio emission predominantly originates from the reverse shock, which means that it is not sensitive to the the structure of the magnetic field upstream of the forward shock. A dedicated study of polarisation of the X-ray emission would be extremely important to solve some of these discrepancies.

Unexpected radio brightening was also observed in the northern region. \citet{Uchiyama:2007} has previously observed localised brightening (and fading; explained by synchrotron cooling) of X-ray emission in RX\,J1713.7$-$3946 on a one year timescale. This was said to be synchrotron emission from electrons quickly accelerated by the diffusive shock acceleration (DSA) process, and the high energy of the variable emission indicated magnetic field amplification in that region of the SNR shock, which corresponds to dense molecular clumps \citep{Sano:2010,Maxted:2013rxj,Sano:2015}. Therefore, the largest brightening increase in the northern parts of \g1 (Figure~\ref{fig:RadioBrightening}) may indicate regions of shock interaction with a highly inhomogeneous ISM. This motivates future X-ray studies probing small-scale X-ray features as well as future ISM observations at arc-sec resolution \citep[e.g. using ALMA, e.g. see][or ATCA]{Sano:2019} to identify shock/ISM interactions.

In addition to exhibiting short-timescale X-ray brightening linked to particle acceleration, SNR RX\,J1713.7-3946 is a strong TeV gamma-ray source \citep{Aharonian:2007rxj,Abdalla:2018rxj} and has been the subject of a number of investigations that search for signatures of CR hadrons accelerated within the SNR shell \citep[e.g.][]{Fukui:2012}. The localised radio brightening of G1.9+0.3 discovered in our study highlights the potential of G1.9+0.3 as a powerful particle accelerator as well. Indeed, as argued in Section~\ref{section:polarisation}, the B-field amplification implied by strong synchrotron X-ray emission can be naturally explained by CRs (e.g. CR streaming instabilities). It follows that this object is a key target in CR origin studies with future high-sensitivity TeV gamma-ray observations of the Cherenkov Telescope Array \citep[CTA, see][]{Acharya:2019}.

\section {Conclusion} \label{section:conclusion}
This radio study of the youngest known Galactic \ac{SNR} has used new \ac{ATCA} observations in 2016 and 2017 and preliminary observations from the Mopra and \ac{MWA} telescopes to examine the flux density, distance, spectral index, polarisation, brightening and expansion of the \g1 shell. Our main findings are:
\begin{itemize}
  \item Our \HI\ and CO study is consistent with a distance of 8.5~kpc;
  \item \g1\ has a mean expansion (in radio continuum) rate of (0.78 $\pm$ 0.09)~per~cent~per~yr over a 31 year period, equivalent to (8800$\pm$1200)~km~s$^{-1}$ at a distance of 8.5\,kpc;
  \item There are very different RM values towards the two so called (east \& west) ``ears'' of \g1. The expansion into the stellar wind of a RGS star progenitor could potentially explain the polarisation characteristics we observe for \ac{SNR} \g1. This hypothesis, however, would have difficulties explaining hydrodynamic properties of the remnant and its observed X-ray emission;
  \item \g1 has a \change{global} spectral index of (-0.81 $\pm$ 0.02), which steepens to $\sim -1$ towards the remnant's north, where we also find increased brightening of up to 195~per~cent, consistent with brightening due to expansion into an inhomogeneous ISM;
  \item There is an average radio brightening of (1.67 $\pm$ 0.35)~per~cent~per~yr;
\end{itemize}

\section*{Acknowledgements}

The \ac{ATCA} is part of the Australia Telescope National Facility which is funded by the Commonwealth of Australia for operation as a National Facility managed by \ac{CSIRO}. This paper includes archived data obtained through the Australia Telescope Online Archive (http://atoa.atnf.csiro.au). We used the \textsc{karma} and \textsc{miriad} software packages developed by the \ac{ATNF} and the \textsc{python} programming language. This work is part of the project 176005 ``Emission nebulae: structure and evolution'' supported by the Ministry of Education, Science, and Technological Development of the Republic of Serbia.

\bibliographystyle{mnras}


\bibliography{G1.9+0.3}                

\begin{thebibliography}{}
\makeatletter
\relax
\def\mn@urlcharsother{\let\do\@makeother \do\$\do\&\do\#\do\^\do\_\do\%\do\~}
\def\mn@doi{\begingroup\mn@urlcharsother \@ifnextchar [ {\mn@doi@}
  {\mn@doi@[]}}
\def\mn@doi@[#1]#2{\def\@tempa{#1}\ifx\@tempa\@empty \href
  {http://dx.doi.org/#2} {doi:#2}\else \href {http://dx.doi.org/#2} {#1}\fi
  \endgroup}
\def\mn@eprint#1#2{\mn@eprint@#1:#2::\@nil}
\def\mn@eprint@arXiv#1{\href {http://arxiv.org/abs/#1} {{\tt arXiv:#1}}}
\def\mn@eprint@dblp#1{\href {http://dblp.uni-trier.de/rec/bibtex/#1.xml}
  {dblp:#1}}
\def\mn@eprint@#1:#2:#3:#4\@nil{\def\@tempa {#1}\def\@tempb {#2}\def\@tempc
  {#3}\ifx \@tempc \@empty \let \@tempc \@tempb \let \@tempb \@tempa \fi \ifx
  \@tempb \@empty \def\@tempb {arXiv}\fi \@ifundefined
  {mn@eprint@\@tempb}{\@tempb:\@tempc}{\expandafter \expandafter \csname
  mn@eprint@\@tempb\endcsname \expandafter{\@tempc}}}

\bibitem[\protect\citeauthoryear{{Aharonian} et~al.,}{{Aharonian}
  et~al.}{2007}]{Aharonian:2007rxj}
{Aharonian} F.,  et~al., 2007, \mn@doi [\aap] {10.1051/0004-6361:20066381},
  464, 235

\bibitem[\protect\citeauthoryear{{Bell}, {Schure}  \& {Reville}}{{Bell}
  et~al.}{2011}]{2011MNRAS.418.1208B}
{Bell} A.~R.,  {Schure} K.~M.,   {Reville} B.,  2011, \mn@doi [\mnras]
  {10.1111/j.1365-2966.2011.19571.x}, \href
  {http://adsabs.harvard.edu/abs/2011MNRAS.418.1208B} {418, 1208}

\bibitem[\protect\citeauthoryear{{Bell}, {Matthews}  \& {Blundell}}{{Bell}
  et~al.}{2019}]{2019MNRAS.488.2466B}
{Bell} A.~R.,  {Matthews} J.~H.,   {Blundell} K.~M.,  2019, \mn@doi [\mnras]
  {10.1093/mnras/stz1805}, \href
  {https://ui.adsabs.harvard.edu/abs/2019MNRAS.488.2466B} {488, 2466}

\bibitem[\protect\citeauthoryear{{Blackwell} et~al.,}{{Blackwell}
  et~al.}{2019}]{Blackwell:2018}
{Blackwell} R.,  et~al., 2019, \pasa, submitted

\bibitem[\protect\citeauthoryear{{Borkowski}, {Reynolds}, {Green}, {Hwang},
  {Petre}, {Krishnamurthy}  \& {Willett}}{{Borkowski}
  et~al.}{2010}]{2010ApJ...724L.161B}
{Borkowski} K.~J.,  {Reynolds} S.~P.,  {Green} D.~A.,  {Hwang} U.,  {Petre} R.,
   {Krishnamurthy} K.,   {Willett} R.,  2010, \mn@doi [\apjl]
  {10.1088/2041-8205/724/2/L161}, \href
  {http://adsabs.harvard.edu/abs/2010ApJ...724L.161B} {724, L161}

\bibitem[\protect\citeauthoryear{{Borkowski}, {Reynolds}, {Hwang}, {Green},
  {Petre}, {Krishnamurthy}  \& {Willett}}{{Borkowski}
  et~al.}{2013}]{2013ApJ...771L...9B}
{Borkowski} K.~J.,  {Reynolds} S.~P.,  {Hwang} U.,  {Green} D.~A.,  {Petre} R.,
   {Krishnamurthy} K.,   {Willett} R.,  2013, \mn@doi [\apjl]
  {10.1088/2041-8205/771/1/L9}, \href
  {http://adsabs.harvard.edu/abs/2013ApJ...771L...9B} {771, L9}

\bibitem[\protect\citeauthoryear{{Borkowski}, {Gwynne}, {Reynolds}, {Green},
  {Hwang}, {Petre}  \& {Willett}}{{Borkowski}
  et~al.}{2017}]{2017ApJ...837L...7B}
{Borkowski} K.~J.,  {Gwynne} P.,  {Reynolds} S.~P.,  {Green} D.~A.,  {Hwang}
  U.,  {Petre} R.,   {Willett} R.,  2017, \mn@doi [\apjl]
  {10.3847/2041-8213/aa618c}, \href
  {http://adsabs.harvard.edu/abs/2017ApJ...837L...7B} {837, L7}

\bibitem[\protect\citeauthoryear{{Bozzetto}, {Filipovi{\'c}}, {Uro{\v
  s}evi{\'c}}, {Kothes}  \& {Crawford}}{{Bozzetto}
  et~al.}{2014}]{2014MNRAS.440.3220B}
{Bozzetto} L.~M.,  {Filipovi{\'c}} M.~D.,  {Uro{\v s}evi{\'c}} D.,  {Kothes}
  R.,   {Crawford} E.~J.,  2014, \mn@doi [\mnras] {10.1093/mnras/stu499}, \href
  {http://adsabs.harvard.edu/abs/2014MNRAS.440.3220B} {440, 3220}

\bibitem[\protect\citeauthoryear{{Bozzetto} et~al.,}{{Bozzetto}
  et~al.}{2017}]{2017ApJS..230....2B}
{Bozzetto} L.~M.,  et~al., 2017, \mn@doi [\apjs] {10.3847/1538-4365/aa653c},
  \href {http://adsabs.harvard.edu/abs/2017ApJS..230....2B} {230, 2}

\bibitem[\protect\citeauthoryear{{Braiding} et~al.,}{{Braiding}
  et~al.}{2018}]{Braiding:2018}
{Braiding} C.,  et~al., 2018, \mn@doi [\pasa] {10.1017/pasa.2018.18}, \href
  {https://ui.adsabs.harvard.edu/abs/2018PASA...35...29B} {35, e029}

\bibitem[\protect\citeauthoryear{{Brose}, {Sushch}, {Pohl}, {Luken},
  {Filipovic}  \& {Lin}}{{Brose} et~al.}{2019}]{2019arXiv190602725B}
{Brose} R.,  {Sushch} I.,  {Pohl} M.,  {Luken} K.~J.,  {Filipovic} M.~D.,
  {Lin} R.,  2019, arXiv e-prints, \href
  {https://ui.adsabs.harvard.edu/abs/2019arXiv190602725B} {p. arXiv:1906.02725}

\bibitem[\protect\citeauthoryear{{Burton} et~al.,}{{Burton}
  et~al.}{2013}]{Burton:2013}
{Burton} M.~G.,  et~al., 2013, \mn@doi [\pasa] {10.1017/pasa.2013.22}, \href
  {http://adsabs.harvard.edu/abs/2013PASA...30...44B} {30, e044}

\bibitem[\protect\citeauthoryear{{Cappellaro}, {Barbon}  \&
  {Turatto}}{{Cappellaro} et~al.}{2005}]{2003astro.ph.10859C}
{Cappellaro} E.,  {Barbon} R.,   {Turatto} M.,  2005, in {Marcaide} J.-M.,
  {Weiler} K.~W.,  eds,  Vol. 99, IAU Colloq. 192: Cosmic Explosions, On the
  10th Anniversary of SN1993J. p.~347, \mn@doi{10.1007/3-540-26633-X\_48}

\bibitem[\protect\citeauthoryear{{Caprioli} \& {Spitkovsky}}{{Caprioli} \&
  {Spitkovsky}}{2014}]{2014ApJ...783...91C}
{Caprioli} D.,  {Spitkovsky} A.,  2014, \mn@doi [\apj]
  {10.1088/0004-637X/783/2/91}, \href
  {https://ui.adsabs.harvard.edu/abs/2014ApJ...783...91C} {783, 91}

\bibitem[\protect\citeauthoryear{{Carlton}, {Borkowski}, {Reynolds}, {Hwang},
  {Petre}, {Green}, {Krishnamurthy}  \& {Willett}}{{Carlton}
  et~al.}{2011}]{2011ApJ...737L..22C}
{Carlton} A.~K.,  {Borkowski} K.~J.,  {Reynolds} S.~P.,  {Hwang} U.,  {Petre}
  R.,  {Green} D.~A.,  {Krishnamurthy} K.,   {Willett} R.,  2011, \mn@doi
  [\apjl] {10.1088/2041-8205/737/1/L22}, \href
  {http://adsabs.harvard.edu/abs/2011ApJ...737L..22C} {737, L22}

\bibitem[\protect\citeauthoryear{{Cherenkov Telescope Array Consortium}
  et~al.,}{{Cherenkov Telescope Array Consortium} et~al.}{2019}]{Acharya:2019}
{Cherenkov Telescope Array Consortium} et~al., 2019, {Science with the
  Cherenkov Telescope Array}.
World Scientific Publishing Co, \mn@doi{10.1142/10986}

\bibitem[\protect\citeauthoryear{{Chevalier}}{{Chevalier}}{1976}]{Chevalier}
{Chevalier} R.~A.,  1976, \mn@doi [\apj] {10.1086/154557}, \href
  {http://adsabs.harvard.edu/abs/1976ApJ...207..872C} {207, 872}

\bibitem[\protect\citeauthoryear{{Cohen}}{{Cohen}}{1975}]{1975MNRAS.171..659C}
{Cohen} R.~J.,  1975, \mn@doi [\mnras] {10.1093/mnras/171.3.659}, \href
  {http://adsabs.harvard.edu/abs/1975MNRAS.171..659C} {171, 659}

\bibitem[\protect\citeauthoryear{{De Horta} et~al.,}{{De Horta}
  et~al.}{2014}]{2014SerAJ.189...41D}
{De Horta} A.~Y.,  et~al., 2014, \mn@doi [Serbian Astronomical Journal]
  {10.2298/SAJ140605001H}, \href
  {http://adsabs.harvard.edu/abs/2014SerAJ.189...41D} {189, 41}

\bibitem[\protect\citeauthoryear{{Dwarkadas}}{{Dwarkadas}}{2005}]{2005ApJ...630..892D}
{Dwarkadas} V.~V.,  2005, \mn@doi [\apj] {10.1086/432109}, \href
  {https://ui.adsabs.harvard.edu/abs/2005ApJ...630..892D} {630, 892}

\bibitem[\protect\citeauthoryear{{Farnes}}{{Farnes}}{2012}]{Farnes:2012PhD}
{Farnes} J.~S.,  2012, PhD thesis, University of Cambridge
  \href{mailto:j.farnes@mrao.cam.ac.uk}{j.farnes@mrao.cam.ac.uk}

\bibitem[\protect\citeauthoryear{{Francis} \& {Anderson}}{{Francis} \&
  {Anderson}}{2014}]{2014MNRAS.441.1105F}
{Francis} C.,  {Anderson} E.,  2014, \mn@doi [\mnras] {10.1093/mnras/stu631},
  \href {https://ui.adsabs.harvard.edu/abs/2014MNRAS.441.1105F} {441, 1105}

\bibitem[\protect\citeauthoryear{{Fukui} et~al.,}{{Fukui}
  et~al.}{2012}]{Fukui:2012}
{Fukui} Y.,  et~al., 2012, \mn@doi [\apj] {10.1088/0004-637X/746/1/82}, \href
  {http://adsabs.harvard.edu/abs/2012ApJ...746...82F} {746, 82}

\bibitem[\protect\citeauthoryear{{G{\'o}mez} \& {Rodr{\'{\i}}guez}}{{G{\'o}mez}
  \& {Rodr{\'{\i}}guez}}{2009}]{Gomez:2009}
{G{\'o}mez} Y.,  {Rodr{\'{\i}}guez} L.~F.,  2009, \rmxaa, \href
  {http://adsabs.harvard.edu/abs/2009RMxAA..45...91G} {45, 91}

\bibitem[\protect\citeauthoryear{{Gooch}}{{Gooch}}{2011}]{kvis}
{Gooch} R.,  2011, {Karma: Visualisation Test-Bed Toolkit}, Astrophysics Source
  Code Library (\mn@eprint {ascl} {1102.018})

\bibitem[\protect\citeauthoryear{{Gray}}{{Gray}}{1994}]{Gray:1994}
{Gray} A.~D.,  1994, \mn@doi [\mnras] {10.1093/mnras/270.4.847}, \href
  {http://adsabs.harvard.edu/abs/1994MNRAS.270..847G} {270, 847}

\bibitem[\protect\citeauthoryear{{Green} \& {Gull}}{{Green} \&
  {Gull}}{1984}]{Green:1984}
{Green} D.~A.,  {Gull} S.~F.,  1984, \mn@doi [\nat] {10.1038/312527a0}, \href
  {http://adsabs.harvard.edu/abs/1984Natur.312..527G} {312, 527}

\bibitem[\protect\citeauthoryear{{Green}, {Reynolds}, {Borkowski}, {Hwang},
  {Harrus}  \& {Petre}}{{Green} et~al.}{2008}]{2008MNRAS.387L..54G}
{Green} D.~A.,  {Reynolds} S.~P.,  {Borkowski} K.~J.,  {Hwang} U.,  {Harrus}
  I.,   {Petre} R.,  2008, \mn@doi [\mnras] {10.1111/j.1745-3933.2008.00484.x},
  \href {http://adsabs.harvard.edu/abs/2008MNRAS.387L..54G} {387, L54}

\bibitem[\protect\citeauthoryear{{Gull}}{{Gull}}{1973}]{1973MNRAS.161...47G}
{Gull} S.~F.,  1973, \mn@doi [\mnras] {10.1093/mnras/161.1.47}, \href
  {http://adsabs.harvard.edu/abs/1973MNRAS.161...47G} {161, 47}

\bibitem[\protect\citeauthoryear{{Gull}}{{Gull}}{1975}]{Gull}
{Gull} S.~F.,  1975, \mn@doi [\mnras] {10.1093/mnras/171.2.263}, \href
  {http://adsabs.harvard.edu/abs/1975MNRAS.171..263G} {171, 263}

\bibitem[\protect\citeauthoryear{{H.E.S.S.~Collaboration}
  et~al.,}{{H.E.S.S.~Collaboration} et~al.}{2018}]{Abdalla:2018rxj}
{H.E.S.S.~Collaboration} et~al., 2018, \mn@doi [\aap]
  {10.1051/0004-6361/201629790}, \href
  {https://ui.adsabs.harvard.edu/abs/2018A%26A...612A...6H} {612, A6}

\bibitem[\protect\citeauthoryear{{Han}, {Manchester}, {van Straten}  \&
  {Demorest}}{{Han} et~al.}{2018}]{2018ApJS..234...11H}
{Han} J.~L.,  {Manchester} R.~N.,  {van Straten} W.,   {Demorest} P.,  2018,
  \mn@doi [\apjs] {10.3847/1538-4365/aa9c45}, \href
  {http://adsabs.harvard.edu/abs/2018ApJS..234...11H} {234, 11}

\bibitem[\protect\citeauthoryear{{Harvey-Smith}, {Gaensler}, {Kothes},
  {Townsend}, {Heald}, {Ng}  \& {Green}}{{Harvey-Smith}
  et~al.}{2010}]{2010ApJ...712.1157H}
{Harvey-Smith} L.,  {Gaensler} B.~M.,  {Kothes} R.,  {Townsend} R.,  {Heald}
  G.~H.,  {Ng} C.-Y.,   {Green} A.~J.,  2010, \mn@doi [\apj]
  {10.1088/0004-637X/712/2/1157}, \href
  {http://adsabs.harvard.edu/abs/2010ApJ...712.1157H} {712, 1157}

\bibitem[\protect\citeauthoryear{Hunter}{Hunter}{2007}]{matplotlib}
Hunter J.~D.,  2007, \mn@doi [Computing In Science \& Engineering]
  {10.1109/MCSE.2007.55}, 9, 90

\bibitem[\protect\citeauthoryear{{Hurley-Walker} et~al.,}{{Hurley-Walker}
  et~al.}{2017}]{2017MNRAS.464.1146H}
{Hurley-Walker} N.,  et~al., 2017, \mn@doi [\mnras] {10.1093/mnras/stw2337},
  \href {http://adsabs.harvard.edu/abs/2017MNRAS.464.1146H} {464, 1146}

\bibitem[\protect\citeauthoryear{{Jiang}, {Zhang}  \& {Fang}}{{Jiang}
  et~al.}{2013}]{2013MNRAS.433.1271J}
{Jiang} Z.~J.,  {Zhang} L.,   {Fang} J.,  2013, \mn@doi [\mnras]
  {10.1093/mnras/stt803}, \href
  {http://cdsads.u-strasbg.fr/abs/2013MNRAS.433.1271J} {433, 1271}

\bibitem[\protect\citeauthoryear{{Johnston}, {McClure-Griffiths}  \&
  {Koribalski}}{{Johnston} et~al.}{2004}]{2004MNRAS.348L..19J}
{Johnston} S.,  {McClure-Griffiths} N.~M.,   {Koribalski} B.,  2004, \mn@doi
  [\mnras] {10.1111/j.1365-2966.2004.07526.x}, \href
  {http://adsabs.harvard.edu/abs/2004MNRAS.348L..19J} {348, L19}

\bibitem[\protect\citeauthoryear{{Kerr} \& {Lynden-Bell}}{{Kerr} \&
  {Lynden-Bell}}{1986}]{1986MNRAS.221.1023K}
{Kerr} F.~J.,  {Lynden-Bell} D.,  1986, \mn@doi [\mnras]
  {10.1093/mnras/221.4.1023}, \href
  {https://ui.adsabs.harvard.edu/abs/1986MNRAS.221.1023K} {221, 1023}

\bibitem[\protect\citeauthoryear{{LaRosa}, {Kassim}, {Lazio}  \&
  {Hyman}}{{LaRosa} et~al.}{2000}]{LaRosa:2000}
{LaRosa} T.~N.,  {Kassim} N.~E.,  {Lazio} T.~J.~W.,   {Hyman} S.~D.,  2000,
  \mn@doi [\aj] {10.1086/301168}, \href
  {http://adsabs.harvard.edu/abs/2000AJ....119..207L} {119, 207}

\bibitem[\protect\citeauthoryear{{Maggi} et~al.,}{{Maggi}
  et~al.}{2019}]{2019arXiv190811234M}
{Maggi} P.,  et~al., 2019, arXiv e-prints, \href
  {https://ui.adsabs.harvard.edu/abs/2019arXiv190811234M} {p. arXiv:1908.11234}

\bibitem[\protect\citeauthoryear{{Manchester}, {Hobbs}, {Teoh}  \&
  {Hobbs}}{{Manchester} et~al.}{2005}]{2005AJ....129.1993M}
{Manchester} R.~N.,  {Hobbs} G.~B.,  {Teoh} A.,   {Hobbs} M.,  2005, \mn@doi
  [\aj] {10.1086/428488}, \href
  {http://adsabs.harvard.edu/abs/2005AJ....129.1993M} {129, 1993}

\bibitem[\protect\citeauthoryear{{Maxted} et~al.,}{{Maxted}
  et~al.}{2013}]{Maxted:2013rxj}
{Maxted} N.,  et~al., 2013, \mn@doi [\pasa] {10.1017/pasa.2013.35}, \href
  {http://adsabs.harvard.edu/abs/2013PASA...30...55M} {30, e055}

\bibitem[\protect\citeauthoryear{{Milne}}{{Milne}}{1987}]{1987AuJPh..40..771M}
{Milne} D.~K.,  1987, \mn@doi [Australian Journal of Physics]
  {10.1071/PH870771}, \href {http://adsabs.harvard.edu/abs/1987AuJPh..40..771M}
  {40, 771}

\bibitem[\protect\citeauthoryear{{Murphy}, {Gaensler}  \&
  {Chatterjee}}{{Murphy} et~al.}{2008}]{2008MNRAS.389L..23M}
{Murphy} T.,  {Gaensler} B.~M.,   {Chatterjee} S.,  2008, \mn@doi [\mnras]
  {10.1111/j.1745-3933.2008.00514.x}, \href
  {http://adsabs.harvard.edu/abs/2008MNRAS.389L..23M} {389, L23}

\bibitem[\protect\citeauthoryear{{Nord}, {Lazio}, {Kassim}, {Hyman}, {LaRosa},
  {Brogan}  \& {Duric}}{{Nord} et~al.}{2004}]{Nord:2004}
{Nord} M.~E.,  {Lazio} T.~J.~W.,  {Kassim} N.~E.,  {Hyman} S.~D.,  {LaRosa}
  T.~N.,  {Brogan} C.~L.,   {Duric} N.,  2004, \mn@doi [\aj] {10.1086/424001},
  \href {http://adsabs.harvard.edu/abs/2004AJ....128.1646N} {128, 1646}

\bibitem[\protect\citeauthoryear{{Pavlovi{\'c}}}{{Pavlovi{\'c}}}{2017}]{2017MNRAS.468.1616P}
{Pavlovi{\'c}} M.~Z.,  2017, \mn@doi [\mnras] {10.1093/mnras/stx497}, \href
  {http://adsabs.harvard.edu/abs/2017MNRAS.468.1616P} {468, 1616}

\bibitem[\protect\citeauthoryear{{Reynolds}, {Borkowski}, {Green}, {Hwang},
  {Harrus}  \& {Petre}}{{Reynolds} et~al.}{2008}]{2008ApJ...680L..41R}
{Reynolds} S.~P.,  {Borkowski} K.~J.,  {Green} D.~A.,  {Hwang} U.,  {Harrus}
  I.,   {Petre} R.,  2008, \mn@doi [\apjl] {10.1086/589570}, \href
  {http://adsabs.harvard.edu/abs/2008ApJ...680L..41R} {680, L41}

\bibitem[\protect\citeauthoryear{{Reynolds}, {Borkowski}, {Green}, {Hwang},
  {Harrus}  \& {Petre}}{{Reynolds} et~al.}{2009}]{2009ApJ...695L.149R}
{Reynolds} S.~P.,  {Borkowski} K.~J.,  {Green} D.~A.,  {Hwang} U.,  {Harrus}
  I.,   {Petre} R.,  2009, \mn@doi [\apjl] {10.1088/0004-637X/695/2/L149},
  \href {http://adsabs.harvard.edu/abs/2009ApJ...695L.149R} {695, L149}

\bibitem[\protect\citeauthoryear{{Reynolds}, {Gaensler}  \&
  {Bocchino}}{{Reynolds} et~al.}{2012}]{2012SSRv..166..231R}
{Reynolds} S.~P.,  {Gaensler} B.~M.,   {Bocchino} F.,  2012, \mn@doi [\ssr]
  {10.1007/s11214-011-9775-y}, \href
  {https://ui.adsabs.harvard.edu/abs/2012SSRv..166..231R} {166, 231}

\bibitem[\protect\citeauthoryear{{Roper} et~al.,}{{Roper}
  et~al.}{2018}]{Roper:2018}
{Roper} Q.,  et~al., 2018, \mn@doi [\mnras] {10.1093/mnras/sty1196}, \href
  {http://adsabs.harvard.edu/abs/2018MNRAS.479.1800R} {479, 1800}

\bibitem[\protect\citeauthoryear{{Roy} \& {Pal}}{{Roy} \&
  {Pal}}{2014}]{2014IAUS..296..197R}
{Roy} S.,  {Pal} S.,  2014, in {Ray} A.,  {McCray} R.~A.,  eds,  IAU Symposium
  Vol. 296, Supernova Environmental Impacts. pp 197--201,
  \mn@doi{10.1017/S1743921313009460}

\bibitem[\protect\citeauthoryear{{Sano} et~al.,}{{Sano}
  et~al.}{2010}]{Sano:2010}
{Sano} H.,  et~al., 2010, \mn@doi [\apj] {10.1088/0004-637X/724/1/59}, \href
  {http://adsabs.harvard.edu/abs/2010ApJ...724...59S} {724, 59}

\bibitem[\protect\citeauthoryear{{Sano} et~al.,}{{Sano}
  et~al.}{2015}]{Sano:2015}
{Sano} H.,  et~al., 2015, \mn@doi [\apj] {10.1088/0004-637X/799/2/175}, \href
  {https://ui.adsabs.harvard.edu/abs/2015ApJ...799..175S} {799, 175}

\bibitem[\protect\citeauthoryear{{Sano} et~al.,}{{Sano}
  et~al.}{2019}]{Sano:2019}
{Sano} H.,  et~al., 2019, \mn@doi [\apj] {10.3847/1538-4357/ab02fd}, \href
  {https://ui.adsabs.harvard.edu/abs/2019ApJ...873...40S} {873, 40}

\bibitem[\protect\citeauthoryear{Sault, Teuben  \& Wright}{Sault
  et~al.}{1995}]{miriad}
Sault R.~J.,  Teuben P.~J.,   Wright M.~C.,  1995, in Astronomical Data
  Analysis Software and Systems IV. p.~433

\bibitem[\protect\citeauthoryear{{Telezhinsky}, {Dwarkadas}  \&
  {Pohl}}{{Telezhinsky} et~al.}{2013}]{2013A&A...552A.102T}
{Telezhinsky} I.,  {Dwarkadas} V.~V.,   {Pohl} M.,  2013, \mn@doi [\aap]
  {10.1051/0004-6361/201220740}, \href
  {http://adsabs.harvard.edu/abs/2013A%26A...552A.102T} {552, A102}

\bibitem[\protect\citeauthoryear{{Uchiyama}, {Aharonian}, {Tanaka}, {Takahashi}
   \& {Maeda}}{{Uchiyama} et~al.}{2007}]{Uchiyama:2007}
{Uchiyama} Y.,  {Aharonian} F.~A.,  {Tanaka} T.,  {Takahashi} T.,   {Maeda} Y.,
   2007, \mn@doi [\nat] {10.1038/nature06210}, \href
  {http://adsabs.harvard.edu/abs/2007Natur.449..576U} {449, 576}

\bibitem[\protect\citeauthoryear{{Uro{\v s}evi{\'c}}}{{Uro{\v
  s}evi{\'c}}}{2014}]{2014Ap&SS.354..541U}
{Uro{\v s}evi{\'c}} D.,  2014, \mn@doi [\apss] {10.1007/s10509-014-2095-4},
  \href {http://cdsads.u-strasbg.fr/abs/2014Ap%26SS.354..541U} {354, 541}

\bibitem[\protect\citeauthoryear{{Uro{\v s}evi{\'c}}, {Pavlovi{\'c}}  \&
  {Arbutina}}{{Uro{\v s}evi{\'c}} et~al.}{2018}]{2018ApJ...855...59U}
{Uro{\v s}evi{\'c}} D.,  {Pavlovi{\'c}} M.~Z.,   {Arbutina} B.,  2018, \mn@doi
  [\apj] {10.3847/1538-4357/aaac2d}, \href
  {http://adsabs.harvard.edu/abs/2018ApJ...855...59U} {855, 59}

\bibitem[\protect\citeauthoryear{{V{\"o}lk}, {Berezhko}  \&
  {Ksenofontov}}{{V{\"o}lk} et~al.}{2003}]{2003A&A...409..563V}
{V{\"o}lk} H.~J.,  {Berezhko} E.~G.,   {Ksenofontov} L.~T.,  2003, \mn@doi
  [\aap] {10.1051/0004-6361:20031082}, \href
  {https://ui.adsabs.harvard.edu/abs/2003A&A...409..563V} {409, 563}

\bibitem[\protect\citeauthoryear{{Wayth} et~al.,}{{Wayth}
  et~al.}{2015}]{GLEAM:2015}
{Wayth} R.~B.,  et~al., 2015, \mn@doi [\pasa] {10.1017/pasa.2015.26}, \href
  {https://ui.adsabs.harvard.edu/abs/2015PASA...32...25W} {32, e025}

\bibitem[\protect\citeauthoryear{{Wilson} et~al.,}{{Wilson}
  et~al.}{2011}]{CABB}
{Wilson} W.~E.,  et~al., 2011, \mn@doi [\mnras]
  {10.1111/j.1365-2966.2011.19054.x}, \href
  {http://adsabs.harvard.edu/abs/2011MNRAS.416..832W} {416, 832}

\bibitem[\protect\citeauthoryear{{Xu} \& {Han}}{{Xu} \&
  {Han}}{2014}]{2014RAA....14..942X}
{Xu} J.,  {Han} J.-L.,  2014, \mn@doi [Research in Astronomy and Astrophysics]
  {10.1088/1674-4527/14/8/005}, \href
  {http://adsabs.harvard.edu/abs/2014RAA....14..942X} {14, 942}

\bibitem[\protect\citeauthoryear{{Yao}, {Manchester}  \& {Wang}}{{Yao}
  et~al.}{2017}]{2017ApJ...835...29Y}
{Yao} J.~M.,  {Manchester} R.~N.,   {Wang} N.,  2017, \mn@doi [\apj]
  {10.3847/1538-4357/835/1/29}, \href
  {http://adsabs.harvard.edu/abs/2017ApJ...835...29Y} {835, 29}

\bibitem[\protect\citeauthoryear{{de Grijs} \& {Bono}}{{de Grijs} \&
  {Bono}}{2016}]{2016ApJS..227....5D}
{de Grijs} R.,  {Bono} G.,  2016, \mn@doi [\apjs] {10.3847/0067-0049/227/1/5},
  \href {https://ui.adsabs.harvard.edu/abs/2016ApJS..227....5D} {227, 5}

\bibitem[\protect\citeauthoryear{{van den Bergh} \& {Tammann}}{{van den Bergh}
  \& {Tammann}}{1991}]{1991ARA&A..29..363V}
{van den Bergh} S.,  {Tammann} G.~A.,  1991, \mn@doi [\araa]
  {10.1146/annurev.aa.29.090191.002051}, \href
  {http://adsabs.harvard.edu/abs/1991ARA%26A..29..363V} {29, 363}

\makeatother
\end{thebibliography}


\bsp	
\label{lastpage}
\end{document}